# Multilayer graphene shows intrinsic resistance peaks in the carrier density dependence.

Taiki Hirahara[1,] Ryoya Ebisuoka[1], Takushi Oka[1], Tomoaki Nakasuga[1], Shingo Tajima[1], Kenji Watanabe[2], Takashi Taniguchi[2] and Ryuta Yagi[1*]

[1] Graduate School of Advanced Sciences of Matter, Hiroshima University,
1-3-1, Kagamiyama Higashi-Hiroshima, Hiroshima 739-8530, Japan.
[2] National Institute for Materials Science (NIMS), 1-1-1, Namiki, Tsukuba, Ibaraki 305-0044, Japan.

Since the advent of graphene, a variety of studies have been performed to elucidate its fundamental physics, or to explore its practical applications. Gate-tunable resistance is one of the most important properties of graphene and has been studied in 1-3 layer graphene in a number of efforts to control the band gap to obtain a large on-off ratio. On the other hand, the transport property of multilayer graphene with more than three layers is less well understood. Here we show a new aspect of multilayer graphene. We found that four-layer graphene shows intrinsic peak structures in the gate voltage dependence of its resistance at zero magnetic field. Measurement of quantum oscillations in magnetic field confirmed that the peaks originate from the specific band structure of graphene and appear at the carrier density for the bottoms of conduction bands and valence bands. The intrinsic peak structures should generally be observed in AB-stacked multilayer graphene. The present results would be significant for understanding the physics of graphene and making graphene FET devices.

*Corresponding author: yagi@hiroshima-u.ac.jp

Introduction

Graphene is a material with excellent physical properties—its electrical mobility, for example, much higher than that of organic semiconductors [1-2]—and is a candidate base material for next-generation devices. Controllability of resistance by using gate voltage is one of most significant properties of graphene for practical application in FET devices. It is known that transport property of graphene is highly influenced by the surface environment (*e.g.*, ionized impurity, contamination or ripples, *etc.* [3-4]). Gate-voltage dependence of the resistance of such dirty graphene samples often shows multiple-peak structure, which possibly originate from local potential energy differences due to the surface contamination. Placing graphene on high-quality *h*-BN flakes or using encapsulation techniques drastically improve the electron mobility of graphene samples and enables one to measure intrinsic features of the transport property of graphene [5-7]. In high-mobility graphene, a large resistance peak appears at the charge neutrality point and resistance decreases monotonically with increasing carrier density [5-7]. This would be a natural property of graphene with a single band, such as mono- and bilayer graphene. Although AB-stacked trilayer graphene has a bilayer-like band and a monolayer band, the gate voltage dependence of its resistance would be similar to that of the single-band system because the density of states of the bilayer-like band is much larger than that of the monolayer band and is dominant [7]. On the other hand, AB-stacked multilayer graphene with larger number of layers does not necessarily show the above mentioned property because it has more than two bilayer-like bands. In this work, we studied AB-stacked tetralayer graphene, which is the simplest case of multilayer graphene with more than three layers. The band structure of AB-stacked tetralayer graphene is shown schematically in panel **i** of Figure 1**a**. The energy of the bottom of the light-mass bilayer-like band is higher than that of the heavy- mass bilayer-like band, and the density of states of the light-mass bilayer-like band is not negligible with respect to that of the heavy-mass bilayer-like band. This paper will show that this multi-bilayer-like band leads to the phenomenon that gate-voltage dependence of resistance generally shows intrinsic peak structures.

Results

Experimental

We employed a technique to modify electronic band structure by using perpendicular electric field. The method was originally proposed to create an energy gap in the graphene system. Monolayer graphene, where electrons behave like massless Dirac fermions with linear dispersion relation [8-10], does not have a band gap. Bilayer graphene, where electron bands have massive dispersion relations [10-12], does not have a band gap either. However, it was predicted that a perpendicular electric field forms an energy gap between the bottoms of conduction and valence bands [13-14]. The response of AB-stacked graphene's electronic structure to perpendicular electric fields shows significant layer-number dependence. AB-stacked trilayer graphene, for example, has a bilayer-like band and monolayer band. Unlike bilayer graphene, it does not exhibit insulating behavior induced by applying a perpendicular electric field. Its electric resistance instead decreases [13-15]. Electronic band structure in AB-stacked tetralayer graphene in the presence of a perpendicular electric field is schematically described in panels **i** to **iii** in Fig. 1**a**. An electric field opens an energy gap at the bottoms of each of the bilayer-like bands as shown in panels **ii** and **iii**. At sufficiently large electric fields, there appears no band overlap arising from semi-metallic features, and the system would become insulating (see panel **iii**).

In order to detect electronic structure by using a transport experiment, the quality of the graphene should be sufficiently high. Moreover, top and bottom electrodes are required to apply a perpendicular electric field. This requirement was attained by using the AB-stacked tetralayer graphene device shown in Figure 1**b**. AB-stacked tetralayer graphene was encapsulated with *h*-BN [16], and onto the top of the stack was transferred a rather thick graphene flake serving as a top gate electrode. This stack of films was formed on a SiO$_2$/Si substrate, where Si is conducting at low temperatures and served as a bottom gate. Encapsulation of graphene by high-quality *h*-BN flakes drastically increases the electric mobility of graphene. The high-quality and atomically flat surface of the *h*-BN flakes reduces scattering due to ionized impurity and surface roughness [6,16-17]. Mobility of the graphene which was calculated with $\mu = 1/ne\rho$, was more than 40,000 cm$^2$/Vs at the high carrier density regime.

### Gate voltage dependence of resistivity at zero magnetic field

Figure 2**a** shows bottom gate voltage $V_{bg}$ dependence of resistance for different top

gate voltages $V_{tg}$ measured at $T$ = 4.2 K. In the trace with $V_{tg} = 0$ V, a slightly complicated peak structures are discernible near the charge neutrality point. The peak structure showed significant variation as top gate voltages was varied. The most prominent variation is enhancement of the peak, which grows with increasing $|V_{tg}|$ values and appeared roughly symmetric with respect to $V_{tg} = 0$ V. The enhancement of the resistance by top gate voltage would be qualitatively understood by formation of a band gap by perpendicular electric fields, as in bilayer graphene [14]. However, the largest peak of $V_{bg}$-dependence of resistivity at $V_{tg} = 0$ V reached to only approximately 250 ohms, and this value was smaller than that in the preceding study in AB-stacked tetralayer graphene that reported insulating behavior [18]. This indicates that energy gap in tetralayer graphene encapsulated with $h$-BN would be considerably small (see the Supplementary Information).

Beside the large peak structure there were small peaks as shown in Fig. 2**a**. These peaks appeared even at $V_{tg} = 0$ V. As shown in the following discussion, the presence of these peaks is an intrinsic property of AB-stacked tetralayer graphene. In the case of dirty graphene, multiple peaks might be observed because inhomogeneity in the sample or surface contaminations can create locally different electrostatic potentials. However, our samples were encapsulated using a dry process, which ensures clean surfaces. The peaks are reproducibly observed in other samples with the same number of layers. Samples with larger numbers of layers also show different peak structures (see Supplementary Information).

In order to study the peak structure further, we measured resistance as a function of the top and the bottom gate voltages (Fig. 2**b**). Gate voltages were converted into carrier densities $n_{tg}$ and $n_{bg}$, which are induced by top and bottom gate voltage, and were calculated by using calibrated specific capacitance. It is clear that the observed peak in Fig. 2**a** showed significant dependence on $n_{tg}$ and $n_{bg}$, and appeared as resistance ridges. By applying both top and bottom gate voltage, total charge density $-en_{tot}$ induced in graphene can be expressed by $-en_{tot} = -e(n_{tg} + n_{bg}) = -(C_{tg}V_{tg} + C_{bg}V_{bg})$ where $C_{tg}$ and $C_{bg}$ are top and bottom gate capacitances, respectively. The white broken line **a** in the figure that passes through $n_{bg} = 0$ and $n_{tg} = 0$ is a resistance ridge that connects two large peaks appearing at $n_{tg} = \pm 0.8 \times 10^{12}$ cm$^{-2}$. On this line, graphene is in a condition of charge neutrality and total carrier density estimated from Hall resistance vanishes. Here, carriers induced by the top gate voltage and bottom gate voltage are compensated, i.e., $C_{tg}V_{tg} + C_{bg}V_{bg} = 0$. The ratio of capacitances, $C_{bg}/C_{tg}$ was calculated to be 0.078. Using calibrated specific bottom gate capacitance, $C_{bg} = 108 \text{aF}/\mu\text{m}^2$, $C_{tg}$ was calculated to be $C_{tg} = 1385 \text{aF}/\mu\text{m}^2$. On the other hand, $D_\perp = -e(n_{tg} - n_{bg}) = C_{tg}V_{tg} - C_{bg}V_{bg}$ is the difference in charge density induced by top and bottom gate voltages and is proportional to the electric flux in graphene. $D_\perp$ is proportional to the electric field perpendicular to the surface of graphene. At the charge neutrality point, ridge **a** appears. In addition, other resistance ridges **b**, **c**, and **d** showed different dependence on gate voltages. In particular ridges **b** and **c** are not parallel to the line of charge neutrality point (ridge **a**), and crossed at approximately $n_{bg} = n_{tg} \approx 0.4 \times 10^{12}$ cm$^{-2}$. The ridge **d** also show a weak dependence as is seen by connecting two largest resistance peaks appearing at $n_{tg} = \pm 0.8 \times 10^{12}$ cm$^{-2}$

In order to inspect the nature of the ridges, we replot it as a function of $n_{tot}$ and electric flux density $D_\perp$ as shown in Fig. 2c. The figure has a few remarkable features. First, ridge structures appear symmetrically about $D_\perp = 0$, from which one could infer that the ridge structures originated from the intrinsic property of graphene. In particular, ridges **b** and **c** overlap at $D_\perp = 0$, forming a single peak. These ridges possibly originated from bottoms of a conduction band and a valence band bottom, of light-mass bilayer-like band. At $D_\perp = 0$, the conduction band and valence band of the light-mass bilayer-like band contact at their bottoms (see left panel **i** in Fig. 2c). The structure of the crossing is reminiscent of the energy gap in the bilayer-like band formed by applying a perpendicular electric field as shown in panel **ii** in Fig. 2c and is possibly related to the light-mass bilayer-like band. Secondly, ridge **d** seems to connect large peaks appearing at $D_\perp \approx \pm 1.5 \times 10^{-7}$ cm$^{-2}$sA and merge into ridge **a**. The growing resistivity with increasing $|D_\perp|$ at $n_{tot} = 0$ would result from decreasing semi-metallic carriers. Density of states for electron-like band and hole-like band tends to decrease. Possibly a small energy gap is formed as shown in inset **iii** of Fig. 2c (see the Supplementary Information).

### Landau level spectroscopy

Energy band structure can be directly probed by quantum oscillation measurement in magnetic fields at $T = 4.2$ K. The structure of Landau levels revealed the nature of the resistance ridges at zero magnetic field. Figure 3**a** displays a map of $R_{xx}$ as a function of $n_{tot}$ and *B*. Here $V_{tg}$ and $V_{bg}$ were varied so that the condition $D_\perp = 0$ was satisfied. Figure 3**b** shows a color map of longitudinal conductivity $\sigma_{xx}$ which was calculated by using a formula, $\sigma_{xx} = \rho_{xx}/(\rho_{xx}^2 + \rho_{xy}^2)$. Before discussing the effect of perpendicular electric fields, here, we briefly explain the fan diagram. Complicated stripes are due to Shubnikov-de Haas oscillation which originated from Landau quantization of two-dimensional electrons.  Each stripe is a Landau level of a specific band, and has a specific index. Landau levels of AB-stacked tetralayer graphene basically consist of two sets of Landau levels for bilayer-like bands: one for the light-mass bilayer-like band and the other for the heavy-mass bilayer-like band. Mixing and crossing of these Landau levels make the structure of the fan diagram complicated [18-19].  Next, we focus on zero-mode Landau levels that appear near the bottoms of bands, and energy gap appearing in the vicinity of charge neutrality points.  Arrows $\beta$ and $\gamma$, which grow vertically with increasing magnetic field, are zero-mode Landau levels of the light-mass and the heavy-mass bilayer-like band. These zero-mode Landau levels generally appear at the bottoms of bands in the Dirac fermions [9]. Analysis of filling factor $\nu$ of energy gaps associated with the Landau level crossing indicated that the zero-mode Landau levels have a degeneracy of eight, which is a hallmark of bilayer-like bands.

Now we discuss the relationship between the fan diagram and the ridge structure at zero magnetic field.  Ridge structures appear approximately the same values of $n_{tot}$ with those of zero-mode Landau levels. Because zero-mode Landau level is formed near the bottoms of the bands, $n_{tot}$ at which the zero-mode Landau level appears would be a measure of $n_{tot}$ for the bottoms of the bands at zero magnetic field. For $D_\perp = 0$, the zero-mode Landau level $\beta$ of light mass bilayer-like band appeared at approximately the same $n_{tot}$ as the crossing point of the ridges **b** and **c** in the lower panel in Fig. 3**a** (the same as Fig. 2**c**).  On the other hand, zero-mode Landau level $\gamma$ for heavy mass bilayer-like band appeared on ridge **d** in the vicinity of $D_\perp = 0$.

 A similar correspondence can be found in the case of non-vanishing vertical electric field. Figures 3**c** and 3**d** show results for $D_\perp = 2.7 \times 10^{-7}$ and $+ 2.7 \times 10^{-7}$ cm$^{-2}$sA, respectively.  Values of $D_\perp$ are indicated by red arrows in lower panels. It is clear that zero-mode Landau level $\beta$ in Fig. 3**a** has split into two zero-modes $\beta_1$ and $\beta_2$ with four-fold degeneracy. ($\beta_2$ also showed small splitting into two levels with two-fold degeneracy at high magnetic fields.)  The zero-mode Landau levels $\beta_1$ and $\beta_2$, appeared at approximately the same $n_{tot}$ of that for ridges **b** and **c** in zero magnetic field (yellow points).  As for the heavy-mass bilayer-like band, splitting is not clearly observed in the fan diagram. However ridge structures similar to **a** and **b** are actually observed for heavy mass bilayer-like band in measurement of wider area using a different sample (see the Supplementary Information).

Ridge **a**, on the other hand, is not relevant to zero-mode Landau levels. As seen in Figs. 3**a** and 3**b**, when at least $B > 0.5$ T, there was an energy gap at around $n_{\text{tot}} = 0$ (indicated with arrow α). This indicates that carrier density is considerably small in zero magnetic field. The ridge **d** in Fig. 3 is influenced by growing resistivity of ridge **a** for $|D_\perp| > \pm\ 0.8 \times 10^{-7} cm^{-2} sA$.

**Behavior of zero-mode Landau levels with respect to $D_\perp$ at a fixed magnetic field.**

The origin of the ridge structure was further checked by detailed measurements of Landau levels as a function of $n_{tot}$ and $D_\perp$. Upper panel in Fig. 4 displays a top and bottom gate voltage dependence of longitudinal resistivity measured in a magnetic field of approximately $B = 3.3$ T. The derivative of longitudinal resistance with respect to total carrier density $n_{tot}$ $(dR_{xx}/dn_{tot})$ is shown to enhance visibility. The step-like structure indicated by $\beta_2$ corresponds to zero-mode Landau level $\beta_2$ in Fig. 3**c**. Two dashed lines of the step shape indicate that Landau level $\beta_2$ split. Zero-mode Landau level $\beta_1$ also split into two levels.  By comparing with lower panel in Fig. 4, which is results for zero magnetic field, it is clear that zero-mode Landau Levels $\beta_1$ and  $\beta_2$ follow the ridge structure **b** and **c**. This clearly indicates that the ridges **b** and **c** originate from bottoms of split light-mass bilayer-like band. The step-like feature originates from crossing of zero-mode with other Landau levels with higher indices. For example, a step appearing at $D_\perp \approx -2 \times 10^{-7}$ cm$^{-2}$sA and $n_{tot}\ =\ 2.7\times\ 10^{12}$ cm$^{-2}$  is a crossing with the Landau level labeled as δ.  On the other hand, the zero-mode Landau level of the heavy-mass bilayer-like band, is not clearly seen in Fig. 4. This is because at the measured magnetic field, the zero-mode Landau level crosses the Landau levels that appear between gaps with filling factors $-12$ and $-8$ at $n_{tot} = -0.4 \times\ 10^{12}$ cm$^{-2}$.

## Discussion
### Origin of resistance ridge structure in zero magnetic field

As seen in the above, the ridges in the carrier density dependence of resistance (or resistance peaks in $n_{tot}$ dependences at fixed values of $D_\perp$) at zero magnetic field reflected electronic band structure. Ridges, which appeared not at the charge neutrality point, are closely related to the bottoms of conduction and valence bands. Conductivity of the samples is contributed by the light-mass bilayer-like band and the heavy-mass bilayer-like band, which are naturally expected to have different mobilities. Therefore, in the frame of two-carrier model conductivity σ is expressed by

$$\sigma = |n_\text{L}|e\mu_L + |n_H|e\mu_H. \tag{1}$$

Here $n_L$ and $n_H$ are carrier densities of the light-mass bilayer-like band and heavy-mass bilayer-like band, and $\mu_L$ and $\mu_H$ are their mobilities. Total carrier density is sum of $n_L$ and $n_H$, i.e., $n_{tot} = n_L + n_H$. The values of the mobility may be different because the band masses of the light-mass bilayer-like band and heavy-mass bilayer-like band should be different. In order to discuss resistance peak structure (or conductance dip structure), one needs to assume $n_{tot}$ dependence of $n_L$ or $n_H$. If we assume a semi-metallic band structure consisting of parabolic bands and assume that $\mu_L$ and $\mu_H$ are constant values, $n_{tot}$ dependence of the conductance is simply given by a polygonal line whose vertices are located at $n_{tot}$ for the bottoms of the bands. This results in discontinuous $d\sigma/dn_{tot}$ at the vertices and generally results in a single dip structure in the conductivity (or a single peak in the resistance). To explain resistance peaks at the bottoms of the bands, the mobility for each band would presumably be dependent on the carrier density. Mobility calculated by using $\mu = \sigma/en_\text{tot}$ generally varies with the carrier density. This phenomenon is widely seen in monolayer and bilayer graphene and results from nature of ionized impurity scattering and screening [20]. Moreover, mobilities would vary at the bottoms of the bands because the band mass would vary near the bottoms of the band owing to mixing of light-mass and heavy-mass bilayer-like bands which is shown in a band calculation considering full Slonczweski-Weiss-McClure (SWMcC) parameters [21] (see the Supplementary

Information).

On the other hand, ridge structure, which was observed at the vicinity of charge neutrality point, would require a bit different explanation. The observed ridge at the charge neutrality point would be related to the report from Ref. [18]: insulating behavior at the charge neutrality point. A small energy gap is possibly formed because of, for example, the staggered potential due to many body interaction of electrons [18,22-23] In this paper we tentatively attributed the enhancement of resistance to possible formation of energy gap (see the Supplementary Information). Electronic states near the charge neutrality point in tetralayer graphene is expected to be complicated as compared to graphene with fewer number of layers, and would require further investigation.

### Spatial symmetry and energy gap

In case of even-layer graphene, spatial inversion symmetry and time reversal symmetry impose strict conditions on the energy state. A conduction band and a valence bands are degenerated at $k = 0$. Therefore, bottoms of conduction and valence bands contact at $k = 0$, and an energy gap does not open [23]. However if spatial inversion symmetry is broken by the potential due to a perpendicular electric field, an energy gap is allowed to open. This fact could be clearly observed by the crossing of ridges **b** and **c** at $D_\perp = 0$. Simple band calculation indicates that gap magnitude generally differs between light-mass and heavy-mass bilayer-like bands, as was observed in the experiment (see Supplementary Information).

### Valley splitting

The broken spatial symmetry caused by the perpendicular electric field is also relevant to the valley splitting of the zero-mode Landau levels. At zero magnetic field, spatial inversion symmetry results in degenerate quantum states for K and K' points in reciprocal lattice space. Quantum states at K and K' points are also degenerated in magnetic fields. The eight-fold degenerated zero-mode Landau levels, which were observed in present experiment for $D_\perp = 0$, are due to inequivalent quantum states for K and K' points.   Zero-mode Landau level splitting due to a perpendicular electric field is a result of breaking of spatial inversion symmetry. Moreover Landau level calculation based on effective mass approximation indicated that the splitting is due to the valley-splitting; quantum states for K and K' points are no longer equivalent, and the zero-mode Landau levels show large valley splitting while those levels with larger indices show smaller valley-splitting, which is consistent with our observation.

In AB-stacked tetralayer graphene samples with a single gate electrode, the valley split zero-mode Landau levels with degeneracy of four or less are always observed rather than eight-fold degenerate Landau levels even though a perpendicular electric field is not intentionally applied by using top and bottom electrodes. This is because spatial inversion symmetry is broken owing to the electrostatic potential due to gate induced charges which screens gate electric fields [24] (see the Supplementary Information).

### Multilayer graphene

The multiple-peak structure in $V_g$ dependence of resistance at zero magnetic field would be a general property of multilayer graphene with four or more layers. In six-layer graphene we found ridge structures similar to those in Figs. 2**b** and 2**c** but with more, which reflected three sets of bilayer-like bands (see the Supplementary Information). These properties of multilayer graphene have not been reported previously. This would be because mobility of the samples was not sufficiently high. Indeed, low mobility samples show a broad peak or broad peaks, and it is difficult to discuss electronic band structure.

### Summary

We have studied transport property of AB-stacked tetralayer graphene samples which are encapsulated with $h$-BN flakes and have top and bottom gate electrodes. We found that, at zero magnetic field, the top and bottom gate voltage dependence of resistance shows non-trivial ridge structures which stem from band structures. By analyzing Landau fan diagrams, positions of some ridge structures were verified to correspond to those of the zero-mode Landau levels of bilayer-like bands. They are eight-fold degenerated at zero perpendicular electric field while they split into two four-fold degenerated levels in presence of perpendicular electric field. Correspondingly, the ridge structures crossed at zero electric field. On the other hand Resistance increased by applying perpendicular electric field at the charge neutrality point at zero magnetic field, however, strong insulating behavior was not observed. In AB-stacked six-layer graphene we also found similar ridge structures. The ridges that appear in the top and bottom gate voltage dependence would be a common feature of multilayer graphene with number of layers larger than three.

## Methods

Graphene was prepared by mechanically exfoliating high-quality Kish graphite using adhesive tape. Graphene was encapsulated with high quality *h*-BN flakes. The top gate was several-layer graphene that was deposited on top of the encapsulated graphene. Each sample was patterned into a Hall bar using standard electron beam lithography. Electric resistance was measured using the standard lock-in technique with a low-frequency excitation current whose frequency and amplitude were typically 10 Hz and 1 µA.  Magnetic fields were applied by using a superconducting solenoid.  Transport measurements were performed at temperature of 4.2 K.

Acknowledgements

This work was supported by KAKENHI No.25107003 from MEXT Japan. Authors thank M. Koshino at Osaka University and R. Saito at Tohoku University for valuable discussions. They also thank Y. Iye at the University of Tokyo for providing high-quality graphite crystals.

## Contributions

TH and RY conceived the experiments. The samples were made by TH, TN, RE, TO and ST. Measurements were carried out by TH and TN. High quality $h$-BN crystals were grown by KW and TT. TH and RY analyzed data and wrote manuscript.

## Competing Interests

The authors declare no conflicts of interest associated with this manuscript.

# Figure Legends

Figure 1 Band structure of tetralayer graphene and sample structure.|
**a** Simplified band structure in tetralayer graphene. **i** Tetralayer graphene consists of a set of light-mass and heavy-mass bilayer-like bands, which are offset in energy. **ii** Perpendicular electric field opens a band gap at the bottoms of the conduction and the valence bands. **iii** Applying a sufficiently large electric field results in forming large gaps and tetralayer graphene becomes insulating. **b** Optical micrograph of encapsulated tetralayer graphene sample that has a top gate (Top left). The scale bar is 10 μm. Schematic structure of an encapsulated graphene stack is shown in the right panel. G is graphene, BN is *h*-BN, Si is Si substrate and SiO$_2$ is SiO$_2$ covering on the Si substrate. Tetralayer graphene was encapsulated with thin *h*-BN flakes, and it was formed on a SiO$_2$/Si substrate. Several layer graphene, which served as a top gate, was deposited onto the top of the encapsulated graphene. Samples were ion-etched into Hall bar shape. Electric contact to the tetralayer graphene was formed by using the technique in Ref. [15]. The lead that is indicated by "Top Gate" in the figure is connected to only the top gate graphene by using the structures illustrated by the two figures at the bottom.

Figure 2  Gate voltage dependence of resistance at zero magnetic field.|
**a** Bottom gate voltage $V_{bg}$ dependence of resistance $R$ for different values of top gate voltages $V_{tg}$. From top to bottom, $V_{tg}$ was varied from 7.5 V to −7.5 V in 2.5 V steps. $T$ = 4.2 K. **b** Map of $R$ as functions of carrier densities $n_{bg}$ and $n_{tg}$, which are tuned by the bottom gate and the top gate voltages and were converted from $V_{tg}$ and $V_{bg}$. Here, $T$ = 4.2 K. $B$ = 0 T. Arrows (a-d) show resistance ridge structure. The white dashed line is the trace of the charge neutrality point. (c) Replot of panel (b) against perpendicular electric flux density D⊥ and carrier density ntot. White broken lines (a-d) show positions of resistance ridges.

Figure 3  Landau fan diagram of tetralayer graphene.|
**a** (Top) Map of longitudinal resistivity $R_{xx}$ at zero perpendicular electric flux density, $D_⊥ = 0$. $T$ = 4.2 K. $β$ and $γ$ indicate position of zero-mode Landau levels of light-mass bilayer-like band and heavy-mass bilayer-like band. α indicates position of energy gap which appears at the charge neutrality point. (Bottom) Resistance ridge structure at zero magnetic field (the same as Fig. 2c) Filled yellow circles indicate positions of the zero-mode Landau levels and charge neutrality point. **b** Map of $\sigma_{xx}$ at $D_\perp = 0$. Filling factors for some gaps are shown. **c** Similar plot as panel a for $D_\perp = -2.7 \times 10^{-7}$ cm$^{-2}$sA. $\beta_1$ and $\beta_2$ indicate split zero-mode Landau levels of light-mass bilayer-like band. $T$= 4.2 K. **d** Similar plot as panel c for $D_\perp = +2.7 \times 10^{-7}$ cm$^{-2}$sA. $T$= 4.2 K.

Figure 4  Top and bottom gate voltage dependences at $B$= 3.3 T and $B$= 0 T. | (Top) A map of derivative of longitudinal resistivity $R_{xx}$ with respect to total carrier density $n_{tot}$ as a function of $n_{tot}$ and $D_\perp$. $B$= 3.3 T and $T$= 4.2 K. Dashed lines show positions of center of some Landau levels. $\beta_1$ and $\beta_2$ denote zero-mode Landau levels of light-mass bilayer-like band. $\delta$ is a Landau level with a higher index. $\delta$ and $\beta_2$ cross at $n_{tot} \approx 1.5 \times 10^{12}$ cm$^{-2}$ and $D_\perp = -2 \times 10^{-7}$ cm$^{-2}$sA. (Bottom) Similar plot of the resistivity at $B$= 0 and $T$= 4.2 K (same as Fig. 2c) Dashed red lines indicate position of zero-mode Landau levels $\beta_1$ and $\beta_2$.

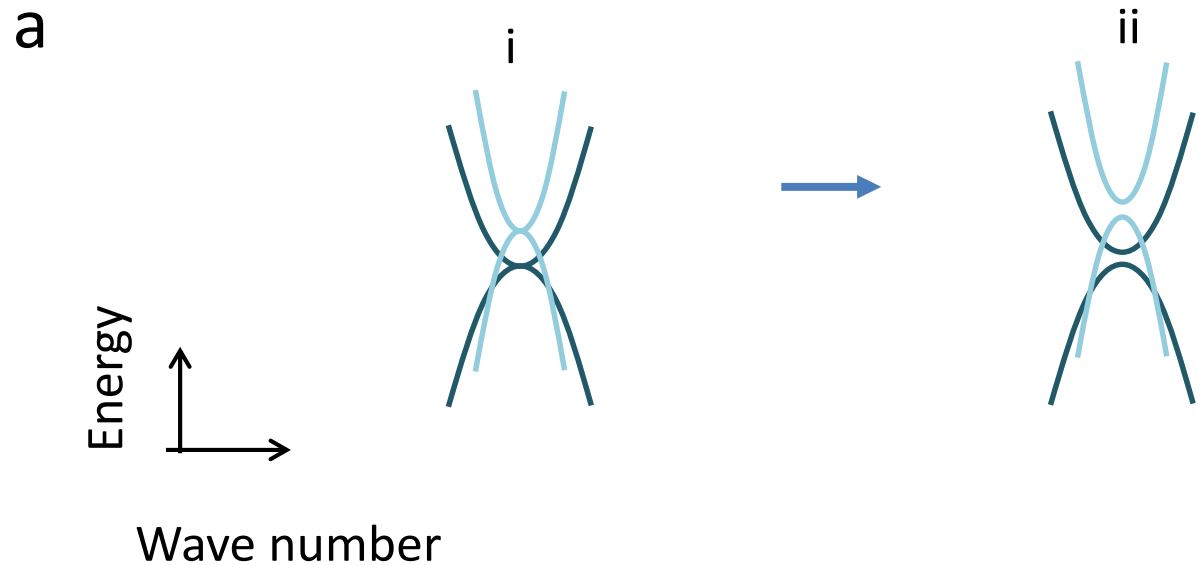

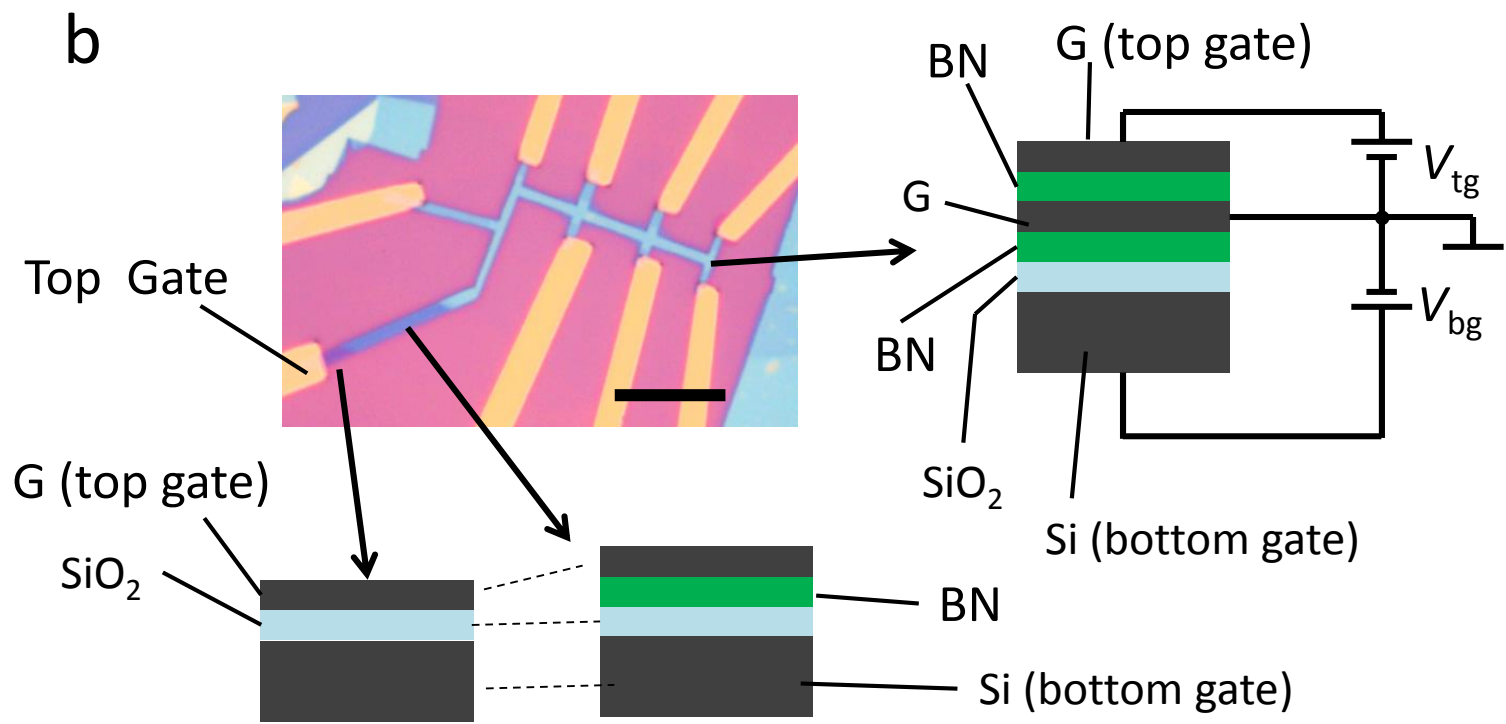

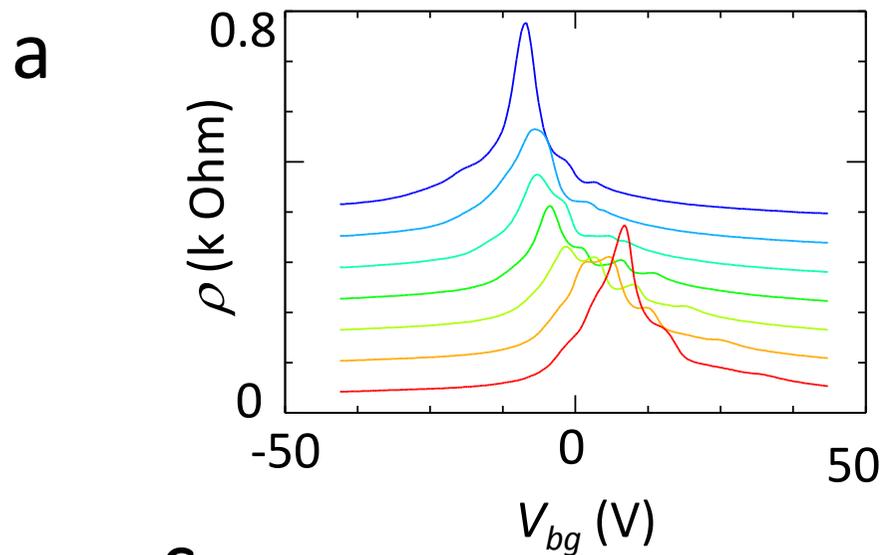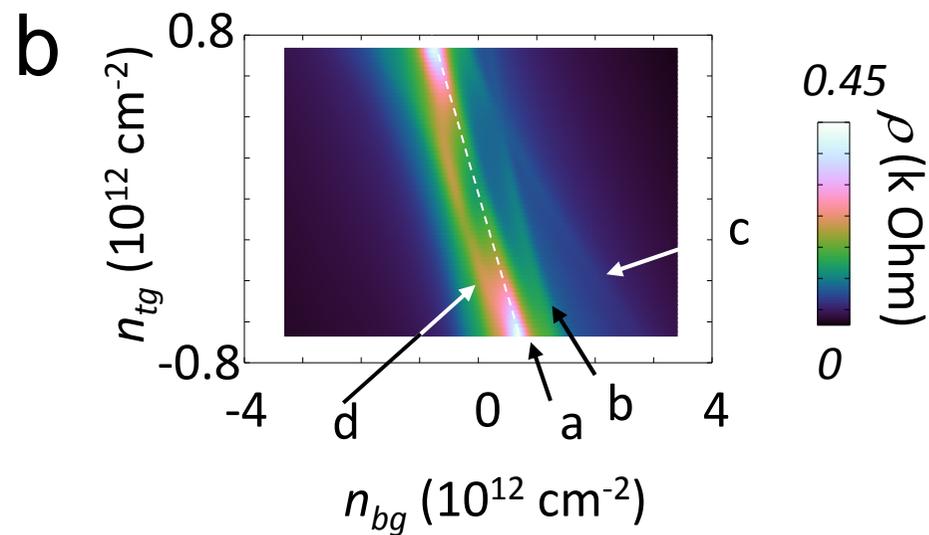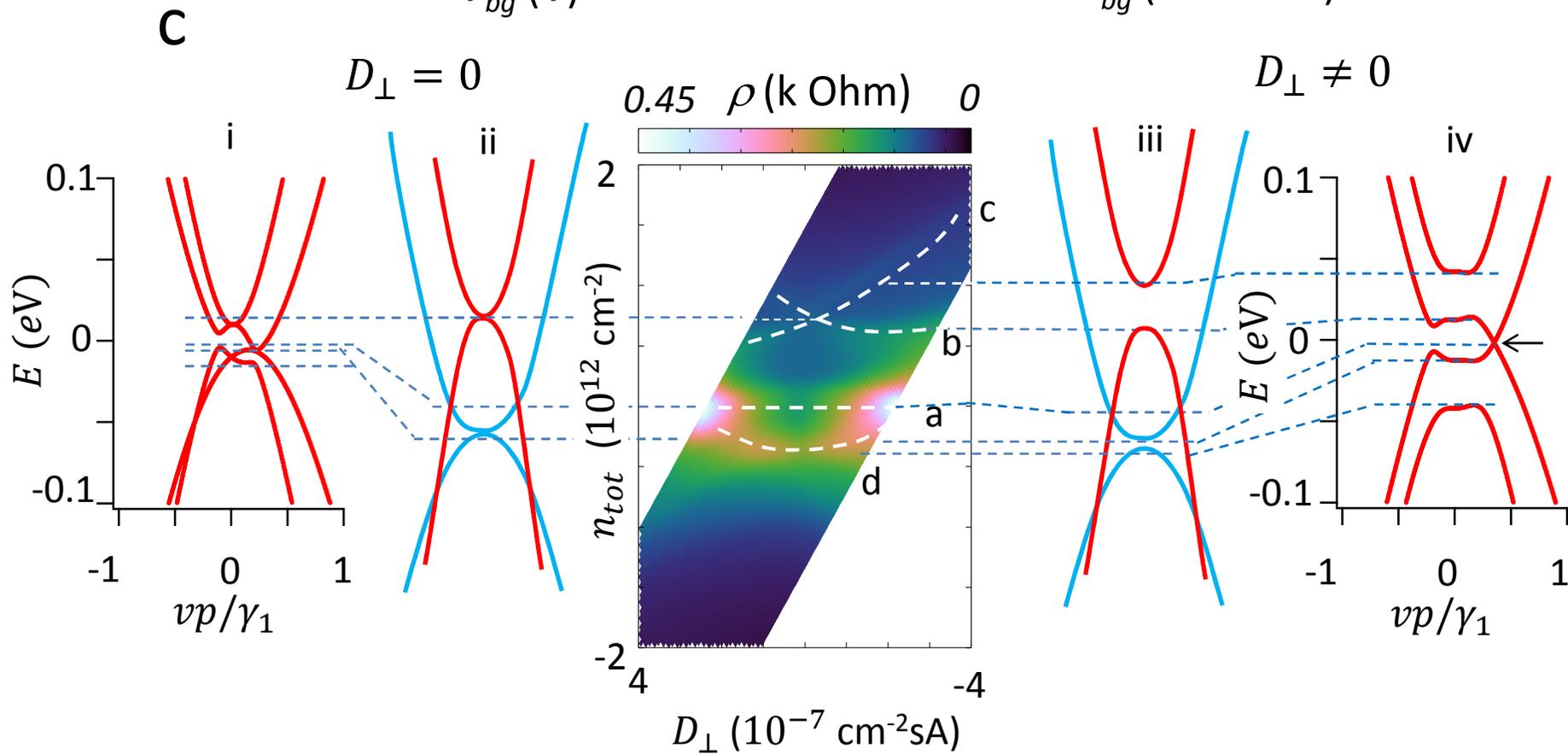

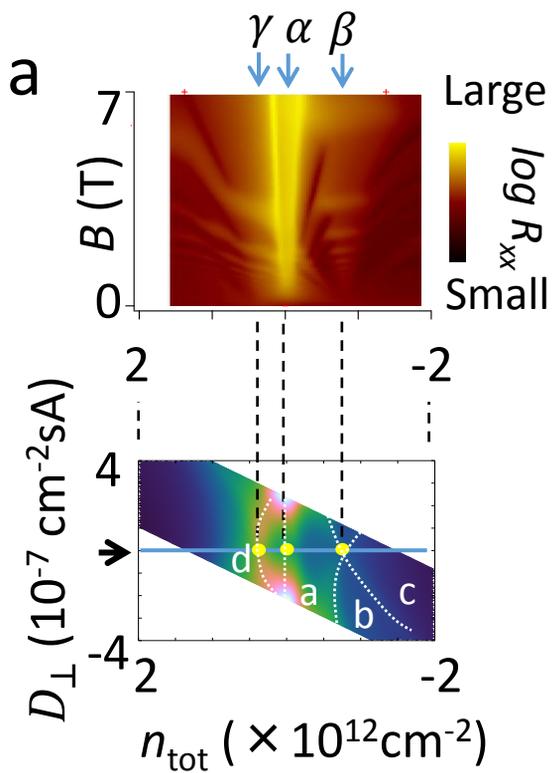
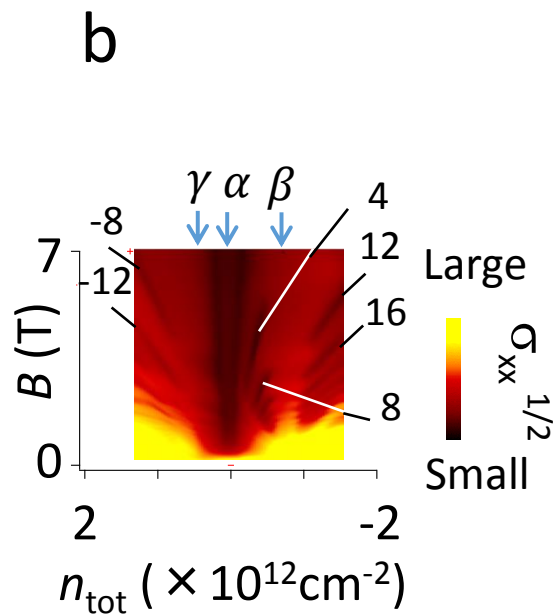
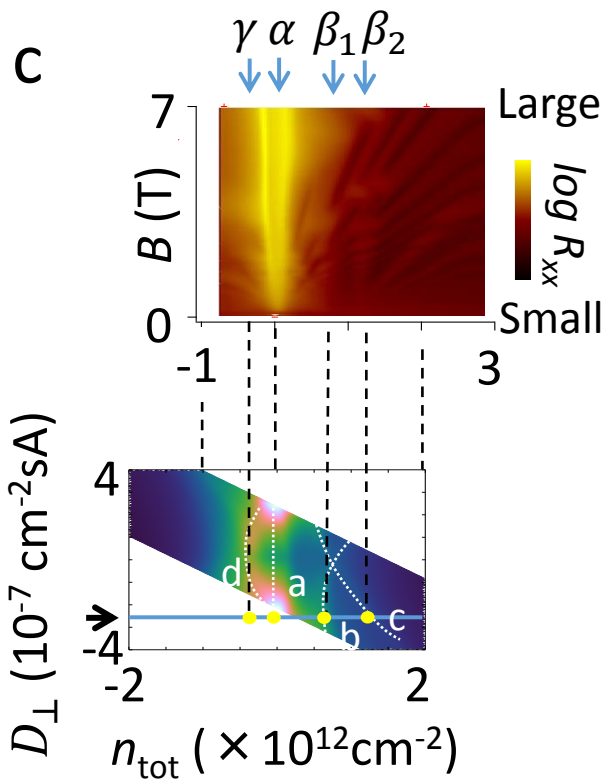
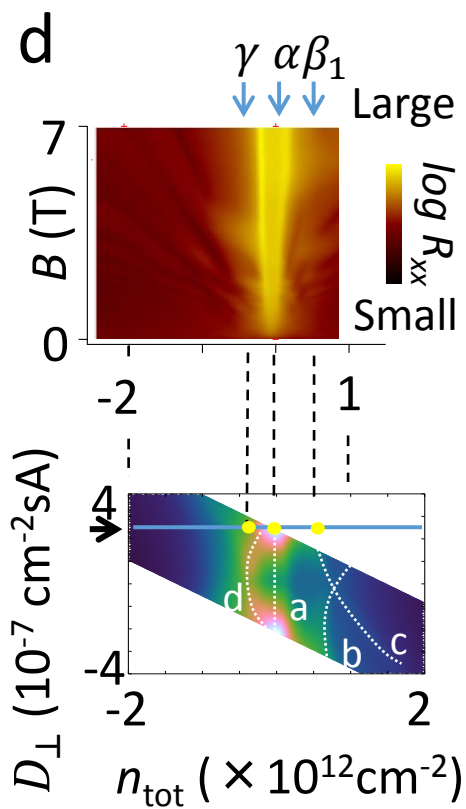

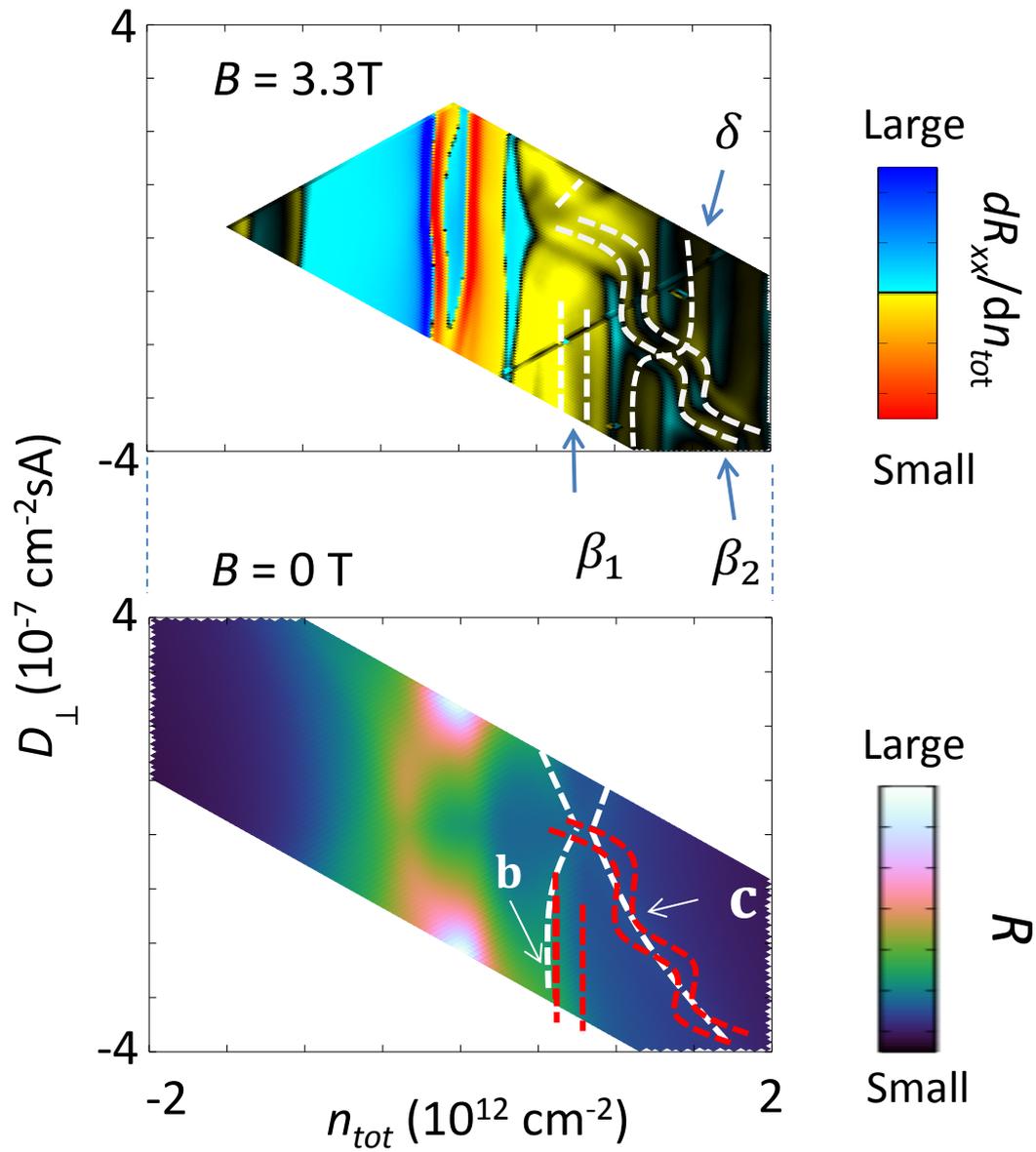

# Supplementary Information

Multilayer graphene shows intrinsic resistance peaks in the carrier density dependence.

Taiki Hirahara[1,] Ryoya Ebisuoka[1], Takushi Oka[1], Tomoaki Nakasuga[1], Shingo Tajima[1], Kenji Watanabe[2], Takashi Taniguchi[2] and Ryuta Yagi[1]

[1] Graduate School of Advanced Sciences of Matter, Hiroshima University, Higashi-Hiroshima, Hiroshima 739-8530, Japan.
[2] National Institute for Materials Science (NIMS), 1-1-1 Namiki, Tsukuba 305-0044, Japan.

### Sample fabrication

Thin flakes of $h$-BN and graphene were made by mechanical cleaving with adhesive tape [1]. Graphene was encapsulated by successively piling up flakes of $h$-BN, graphene and $h$-BN on SiO$_2$ substrates by using a dry-process transferring technique with poly (propylene carbonate) (PPC) films [2]. Several-layer graphene was transferred onto the top of the encapsulated graphene by using the washer method [3]. The sample was patterned by using electron beam lithography (EBL) and plasma etching. The electrical leads were formed using EBL and vacuum deposition of Cr and Au. Electrical contacts between the leads and graphene were attained using the technique described in Ref. [2].

### Mobility of Sample

Figure S1 shows gate voltage dependence of electric mobility that was calculated using $\mu = 1/en_{tot}\rho$. Mobility at $V_{bg} = \pm 50$ V and $V_{tg} = 0$ V is about 40,000 cm²/Vs.

**Checking number of layers and stacking order of tetralayer graphene**

Determinations of number of layers and stacking order are crucial in the study of multilayer graphene. They were determined by color contrast of the digitized optical image, Raman G' spectra shape, AFM topography and the Landau fan diagram. Number of layers can be readily identified by analyzing color digit of the optical images of graphene flakes which were exfoliated on the SiO₂ (300 nm thick )/Si (p+ doped) substrate. First, we calibrated relationship between the color and the thickness. The optical micrograph of a representative sample which was used in the calibration is shown in Fig. S2**a**. We checked the thickness of graphene by using the AFM measurements. An example of AFM topography is shown in Fig. S2**b**, which is measured for the square region in Fig. S2**a**. A line scan along the red line is shown in Fig. S2**c**. Step height is given by about 0.33 nm/ layer, which is approximately the same as the inter-layer distance of graphite.

For many graphene samples with identified number of layers, we performed the Raman spectra measurements. Figure S3**a** shows Raman G' band spectra of graphene flakes with 2 to 5 layers. Here, the data were offset and normalized. These approximately reproduce the results of AB-stacked graphene which have been reported by other groups [8-10]. The spectra shown in Fig. S3**a** appeared most frequently in graphene sample prepared from our Kish graphite crystals. Actually, in particular batches of graphite crystals that was used to prepare the sample for the transport experiment, we did not find any graphene flakes that show G' band spectra shapes other than Fig. S3**a.** In other batches of graphite crystals, we found flakes that show Raman G' band spectral shape for ABC-stacking [9-11], as shown in Fig.

S3**b**.

For the sample that has been shown in the main text, we could not perform the Raman measurement because performing the Raman measurements during the fabrication process might contaminate surfaces of graphene or *h*-BN. In addition, after finishing making sample, the Raman measurement is impossible because graphene to be measured is underneath thick graphene and *h*-BN stack serving as the top gate electrode. However, a proof for the AB-stacked tetralayer graphene can be found in the structure of Landau levels. Fig. S4**a** shows the Landau fan diagram of another AB-stacked tetra-layer graphene sample, which was simply formed on a *h*-BN flake and not encapsulated. This sample was identified to be AB-stacked tetralayer graphene by the Raman spectroscopy. The Raman G' band spectra shape indicated AB-stacked tetralayer graphene. Moreover, a map of Raman G band spectral intensity indicated that the region relevant to the transport experiment is uniform as shown in Fig. S4**b**, which ensures no ABC stacks in the region.

The basic structures of Landau level are approximately the same between the abovementioned sample and the samples described in the main text and supplementary: conspicuous energy gaps and zero-mode Landau levels appeared approximately the same positions as shown with blue lines in Figs. S4**a** and S4**c**, (The data displayed in Fig. S4**c** are the same as Figs. 3**c** and **d** in the main text). The Landau level structure reflects the dispersion relation, and it results from the number of layers and stacking order of graphene. Therefore, data which are displayed in the main text are for AB-stacked tetralayer graphene.

In both Figs. S4**a** and S4 **c**, the zero-mode Landau level of light-mass bilayer-like band which is eight-fold degenerate at $D_\perp = 0$ (Fig. 3**a** in main text) split into two levels with four-fold degeneracy. We found that the gap width in $n_{tot}$ between the split of zero-mode Landau levels is slightly different between Figs. S4**a** and S4**c**. This originates from difference in the effective perpendicular electric fields. In case of Fig. S4**c**, perpendicular electric fields (electric flux density) with a fixed magnitude are applied by using top and bottom gate electrode. On the other hand, in case of Fig. S4, the sample has a single bottom gate electrode, so that magnitude of the perpendicular electric field (electric flux density) varies with gate voltages. Carriers that are induced by gate voltage are expected to distribute in graphene over a characteristic screening length.

## Reproducibility of the ridge structures at zero magnetic field

Figure S5 **a** shows $n_{bg}$ and $n_{tg}$ dependence of resistivity $\rho$, which was measured in a different AB-stacked tetralayer graphene sample (Sample B). Figure S5**b** shows derivative of rho with respect to $n_{bg}$. Fig. S5**c** is a replot of the same data as Fig. 2**b** in the main text (Sample A). A measured region for Fig. S5**c** is indicated with a yellow square in Fig. S5**b**. As seen in the figures, ridges **a**, **b** and **c** were also observed in the both samples. The reproducibility of the ridge structures ensures that the structures originated from the intrinsic nature of tetralayer graphene. In addition, new ridges **e** and **f** are seen in Fig. S5**a** and **b**. The ridges **e** and **f** would arise from band bottoms of the heavy-mass bilayer-like band. Figure S6**a** and **b** show replots of Fig. S5**a** and **c** as a function of $n_{tot}$ and $D_\perp$.

## Results in six-layer graphene

Figure S7**a** and **b** show similar plot as Fig. 3**b** and 3**c** in the main text for a AB-stacked six-layer graphene sample. The resistance ridge structures that vary with $n_{tg}$ and $n_{bg}$ are also seen, as in the case of tetralayer graphene.

Panel **c** shows **a** derivative of resistance with respect to $D_\perp$. The ridge structure is more complicated than that of the tetralayer. Formation of energy gap similar to the light-mass bilayer-like bands in AB-stacked tetralayer graphene is clearly visible as indicated by dashed lines which show a crossing feature. Here, we do not go into detailed band structures which will be reported elsewhere, but zero-mode Landau levels appear at the position of the intrinsic resistance peaks (not shown).

**Band structure of tetralayer graphene**

We calculated the energy gap in AB-stacked tetralayer graphene using the standard Hamiltonian for effective mass approximation [4, 5]. Koshino has shown that Hamiltonian *H* is decomposed into submatrices by a unitary transformation as

$$H = \begin{pmatrix} H_b & H_c \\ H_c & H_B \end{pmatrix},$$

(SF.1)

Here submatirces $H_b$ and $H_B$ are those for the light-mass and heavy-mass bilayer-like bands, respectively, and are given by

$$H_b = \begin{pmatrix} \lambda_+^2 U_1 + \lambda_-^2 U_3 - \lambda\gamma_2 & v\pi^\dagger & -\lambda_b v_4 \pi^\dagger & \lambda_b v_3 \pi \\ v\pi & \lambda_+^2 U_1 + \lambda_-^2 U_3 + \Delta' - \lambda\gamma_5 & \lambda_b \gamma_1 & -\lambda_b v_4 \pi^\dagger \\ -\lambda_b v_4 \pi & \lambda_b \gamma_1 & \lambda_+^2 U_4 + \lambda_-^2 U_2 + \Delta' - \lambda\gamma_5 & v\pi^\dagger \\ \lambda_b v_3 \pi^\dagger & -\lambda_b v_4 \pi & v\pi & \lambda_+^2 U_4 + \lambda_-^2 U_2 - \lambda\gamma_2 \end{pmatrix},$$

(SF.2)

$H_B$

$$= \begin{pmatrix} \lambda_+^2 U_3 + \lambda_-^2 U_1 + \lambda\gamma_2 & v\pi^\dagger & -\lambda_B v_4 \pi^\dagger & \lambda_B v_3 \pi \\ v\pi & \lambda_+^2 U_3 + \lambda_-^2 U_1 + \Delta' + \lambda\gamma_5 & \lambda_B \gamma_1 & -\lambda_B v_4 \pi^\dagger \\ -\lambda_B v_4 \pi & \lambda_B \gamma_1 & \lambda_+^2 U_2 + \lambda_-^2 U_4 + \Delta' + \lambda\gamma_5 & v\pi^\dagger \\ \lambda_B v_3 \pi^\dagger & -\lambda_B v_4 \pi & v\pi & \lambda_+^2 U_2 + \lambda_-^2 U_4 + \lambda\gamma_2 \end{pmatrix}.$$

(SF.3)

Here $\pi = \hbar(k_x + ik_y)$, $v = (\sqrt{3}/2)a\gamma_0/\hbar$, and $v_1 = (\sqrt{3}/2)a\gamma_1/\hbar$, where $a = 0.246$ nm is the lattice constant of the graphene, and $\gamma_i$ ($i = 0,1,2,...,5$) an $\Delta'$ are SWMcC (Slonczewski-Weiss-McClure) parameters [6, 7]. $\lambda_\pm$, $\lambda_b$, $\lambda_B$, $\lambda$ are constants given by

$$\lambda_\pm = \sqrt{\frac{5 \pm \sqrt{5}}{10}}, \tag{SF.4}$$

$$\lambda_b = (-1 + \sqrt{5})/2, \tag{SF.5}$$

$$\lambda_B = (1 + \sqrt{5})/2, \tag{SF.6}$$

$$\lambda = 1/\sqrt{5}. \tag{SF.7}$$

We added to the diagonal element of the Hamiltonian matrix the potential energy for each layer $U_i$, where $i$ is an index of layer number. The submatirix $H_c$, which couples two bilayer-like bands, is given by

$H_c$

$$= \begin{pmatrix} \lambda(U_1 - U_3 + \gamma_2/2) & 0 & 0 & 0 \\ 0 & \lambda(U_1 - U_3 + \gamma_5/2) & 0 & 0 \\ 0 & 0 & \lambda(U_2 - U_4 - \gamma_5/2) & 0 \\ 0 & 0 & 0 & \lambda(U_2 - U_4 - \gamma_2/2) \end{pmatrix},$$

(SF.8)

If $H_c$ is ignored, the energies of the bottoms of the conduction and valence band are analytically calculated to be

$$\varepsilon_{k=0}^{light} = \begin{cases} \lambda_+^2 U_1 + \lambda_-^2 U_3 - p\gamma_2 \\ \lambda_+^2 U_4 + \lambda_-^2 U_2 - p\gamma_2, \end{cases}$$

(SF9)

and

$$\varepsilon_{k=0}^{heavy} = \begin{cases} \lambda_+^2 U_3 + \lambda_-^2 U_1 + p\gamma_2 \\ \lambda_+^2 U_2 + \lambda_-^2 U_4 + p\gamma_2. \end{cases}$$

(SF10)

Therefore, energy gaps at $k = 0$ are simply calculated to be

$$\Delta\varepsilon_{k=0}^{light} = |\lambda_+^2(U_1 - U_4) + \lambda_-^2(U_3 - U_2)|,$$

(SF11)

and

$$\Delta\varepsilon_{k=0}^{heavy} = |\lambda_+^2(U_3 - U_2) + \lambda_-^2(U_1 - U_4)|.$$

(SF12)

Here, we consider a particular case that the potential energy is antisymmetric, *i.e.*, $U_1 = -U_4$ and $U_2 = -U_3$. Then the differences of the potential energy $U_3 - U_2$ and $U_1 - U_4$ can be written by using the two parameters $\Delta U = U_1 - U_4$ and $r$ as

$$U_2 - U_3 = r(U_1 - U_4) = r\Delta U.$$

(SF13)

Then,

$$\Delta\varepsilon_{k=0}^{light} = (\lambda_+^2 - r\lambda_-^2)|\Delta U|,$$

(SF14)

And

$$\Delta\varepsilon_{k=0}^{heavy} = (\lambda_-^2 - r\lambda_+^2)|\Delta U|.$$

(SF15)

In a weak external field, the gaps are approximately proportional to $|D_\perp|$ because $|\Delta U|$ is roughly proportional to $|D_\perp|$. The light-mass bilayer-like band is expected to have a larger energy gap than the heavy-mass bilayer-like band.

**Landau levels in tetralayer graphene.**

Landau levels of tetralayer graphene are valley-degenerated because the system has a spatial inversion symmetry. The spatial inversion symmetry can be broken by applying electric fields by using gate voltages and giving potential variations in each layer. Then the valley degeneracy of the Landau levels should be lifted. Here we show this by a simple numerical calculation of the energy spectra of tetralayer graphene in a magnetic field. We calculated the energy spectra in the frame of effective mass approximation that considers all the SWMcC parameters [4]. We expanded the wave functions with the Landau functions, and the electric potential at each layer was added to the diagonal element of the Hamiltonian matrix. Figure S8**a** shows energy spectra of tetralayer graphene in the absence of external potential variation. The SWMcC parameters of this calculation were taken to be the same as those of graphite, *i.e.*, $\gamma_0 = 3.16$ eV, $\gamma_1 = 0.39$ eV, $\gamma_2 = -0.02$ eV, $\gamma_3 = 0.31$ eV, $\gamma_4 = 0.44$ eV, $\gamma_5 = 0.038$ eV, and $\Delta' = \Delta - \gamma_2 + \gamma_5 = 0.037$ eV. Spectra for the K valley and K' valley are identical so that each line is four-fold degenerated because of the degeneracy in spin and valley degrees of freedom. Next we show the result in the presence of potential variation in graphene. We assumed a model potential. From the top layer to the bottom layer, the electric potential energy was $-0.01$, $-0.005$, $+0.005$, and $+0.01$ eV. This breaks spatial inversion symmetry. Figure S8**b** shows results of the calculation. Red and black lines are energy eigenvalues for the K and the K' valleys, respectively. It is seen that energy spectra of K and K' valleys differ slightly. In particular, zero-mode Landau levels showed splitting, which corresponds to forming an energy gap at the bottoms of the conduction and valence bands of the bilayer-like band. Valley splitting can be seen in the energy spectra of trilayer graphene even in the absence of external potential variation [4].

**On insulating behavior at $D_\perp = 0$**

In four different samples of AB-stacked tetralayer graphene, we have not observed the strong insulating behavior that has been reported in Ref. 18 in the main text. The study attributed the insulating behavior to possible staggered potential. We estimated maximum possible amplitude of the staggered potential at the neutrality point by model calculations. With the increasing strength of the staggered potential, the zero-mode Landau levels split as shown by the arrows in Fig. S9**a**.  In our experiments, splitting of the zero-mode Landau level was not observed for $D_\perp = 0$. (Fig. S9**b**.) This would indicate that the staggered potential is absent or its amplitude is significantly small. A model calculation indicates that if the staggered potential amplitude is less than about ~0.0005 eV, splitting of zero-mode Landau levels is within the width of the zero-mode Landau levels in the experimental fan diagram. Therefore, presence of staggered potential does not contradict with our experiment as long as the amplitude is sufficiently small. However, to be discussed in the next section, energy gap does not form between the conduction and valence bands for this magnitude of staggered potential at $D_\perp = 0$.

The origin of the difference between our results and Ref. 18 in the main text is currently unknown. This might be sought in the experimental setup: graphene is suspended in Ref. 18 in the main text, while it is encapsulated in the present study.

**Dispersion relation in presence of perpendicular electric field.**

Next we discuss the dispersion relation of electronic band structure at zero magnetic fields. Variations of the band structure as a function of a perpendicular electric field were calculated numerically using the effective mass approximation with Slonczewski-Weiss-McClure parameters of graphite. Figure S10 shows results in the presence of electric fields which are created by charges induced by top gate and the bottom gate voltages. Here, $n_t$ and $n_b$ are the charges induced by top and bottom gate voltages, respectively, and $p = \hbar k_x$. We assumed characteristic screening length of 0.43 nm and permittivity of graphene of 2. Figure S11 shows enlargement plots near $E/\gamma_0 = 0$.

In the absence of perpendicular electric field, *i.e.*, $n_t = 0$ and $n_b = 0$, the calculated dispersion relation is essentially the same as that reported in [4, 5]. Tetalayer graphene has two sets of bilayer-like bands. Coupling between these bands tends to open an energy gap in the vicinity of *E* = 0. However, because of the trigonal warping, energy gap closes partially, and conduction band and valence band touches or overlaps at very small regions in *k*-space [4, 5] (Point X in Fig. S10).

The perpendicular electric field varies the dispersion relations, as seen in Figure S10. The most conspicuous change is that energy gaps are formed between band A and B, and between band C and D. Bottoms of the bands are flattened, which increases band masses. In addition, mini-Dirac cones appear near $E/\gamma_0 = 0$. Even with increasing $|D_\perp| = |e(n_b - n_t)|$, the valence band and the conduction band remain touching, as seen in Figs. S10 and S11. The magnitude of group velocity near the touching point is seen to become larger with increasing $|D_\perp|$.

The variation of the dispersion in the vicinity of *E* = 0 could result in considerable change in the resistivity. In addition, the variation of the resistivity could originate from formation of energy gap between conduction and valence band. Indeed, the band structure in the vicinity of $E = 0$ is strongly dependent on $\gamma_3$ in the SWMcC parameters, which describes the trigonal warping. In the case of $\gamma_3 = 0.28$ eV, which is slightly smaller than the parameter of graphite, an energy gap opens between bands B and C in the presence of perpendicular electric field as shown in Figure S12, while the conduction and valence bands are touching or overlapped at $|D_\perp| = 0$.

Even for SWMcC parameter of graphite, opening the energy gap via the perpendicular electric field is possible if the staggered potential is present. Figures S13 and S14 show the results of for $\gamma_3 = 0.3$ eV and with staggered potentials with amplitude of 0.5 and 1 meV, respectively. The perpendicular electric field opens energy gaps with a few mil electron volts between bands B and C, while, at zero electric field, no energy gap open.

The simplified band structure described in the main text would have the characteristic features of the above band models in terms of formation of the energy gap.

## Thickness of *h*-BN and dielectric breakdown

In the sample fabrication process, we have not measured the thickness of the *h*-BN flakes of the top gate insulator by using atomic force microscopy (AFM) to avoid contamination. The thickness of top *h*-BN underneath the top gate graphene is difficult to measure by using AFM after finishing making samples because of the sample structure. We estimated the thickness of the *h*-BN flake by using two different methods other than the AFM measurement. The first method is to analyze the color of the *h*-BN flakes. We calibrated the relationship between the thickness that was measured by AFM and the intensity of the RGB-signal of the digitized optical micrograph of *h*-BN flakes, which were exfoliated on the $SiO_2$/Si substrate. The thicknesses of the *h*-BN flakes for the top gate insulator were estimated to be $15 \pm 4$ nm for sample A (data are presented in the main text), and $65 \pm 4$ nm for sample B (data are presented in Figs. S2a). The thicknesses of the bottom *h*-BN flakes were both about 10 nm for samples A and B.

The second method is to use the ratio of capacitances $C_{tg}/C_{bg}$. According to the classical theory of electromagnetism, $C_{tg} = \epsilon_0 \epsilon_{hBN}/d_{hBN}$ and $C_{bg} = \epsilon_0 \epsilon_{SiO2}/d_{SiO2}$. Here, $d_{hBN}$ and $d_{SiO2}$ are the thicknesses of the gate insulators, i.e, h-BN and SiO$_2$, respectively. $\epsilon_0$ is the dielectric constant of vacuum. $\epsilon_{hBN}$ and $\epsilon_{SiO2}$ are the relative permittivity of h-BN and SiO$_2$, respectively. $\epsilon_{hBN}$ is about 3.9. $\epsilon_{hBN}$ is reported to be 3-4 [12, 13], which is close to the value of $\epsilon_{hBN}$. The ratio $C_{tg}/C_{bg}$ was estimated from the condition of $n_{tot} = 0$ to be 13.4 and 3.82 for samples A and B, respectively. If we ignore the thickness of bottom h-BN (~ 10nm) which is much thinner than that of SiO$_2$ ($d_{SiO2}$= 300nm), $d_{hBN}$ is estimated to be 23 and 89 nm for samples A and B, respectively.

The dielectric breakdown in mechanically exfoliated h-BN is reported to occur at about 7 MV/cm [14-16]. Breakdown voltages for the top gate are estimated to be 16 V and 55 V in samples A and B. Therefore experiments were done well below the breakdown voltages.

What happens when the gate voltage exceeds dielectric breakdown limit? Figure S15 shows an example of the dielectric breakdown in a AB-stacked six-layer graphene sample (the sample is different from that for Fig. S7). In this sample, there was a considerable leak current through the top gate electrode for $|V_{tg}|$>5V, and therefore, the ridge structure appearing at $n_{tot} = 0$ is curved. This clearly indicates that top gate does not operate normally for $|V_{tg}|$>5V, and is strikingly different from the results in the AB-stacked six-layer sample which showed a negligible gate leak current. From these results, we can conclude the effects of dielectric breakdown cannot be observed in the data displayed in the main text, and Figs. S2**a** and **c**.

# Figure Captions

**Figure S1 The mobility of the sample. |**

$V_{bg}$-dependence of mobility $\mu = 1/n_{bg}e\rho$ of the tetralayer graphene sample. $T$ = 4.2 K, and $B$ = 0. $V_{tg}$= 0 V.

**Figure S2 Number of layers for graphene flakes. |**

**a** An optical micrograph of a graphene flake consisting of regions with different thicknesses. Numbers show number of layers. **b** AFM Topography for the region displayed with yellow square in panel a.  **c** A topographic line scan along the line indicated in panel **b**.

**Figure S3 The spectral shapes of Raman G' band for a few layer graphene. |**

**a** Raman spectra of G' band for AB-stacked graphene.  The intensity of the spectra is normalized and offset . The spectral shapes of the G' band varied systematically from 2 to 5 layers. **b** Raman spectra of G' band spectra for ABC-stacked trilayer graphene and ABC-stacked tetralayer graphene.

**Figure S4 The Landau fan diagrams of AB-stacked tetralayer graphene |**

**a** Landau fan diagram of the AB-stacked tetra-layer graphene sample which is different from those described in the text. $T$ = 4.2 K. Graphene is not encapsulated but placed simply on *h*-BN. Some energy gaps between Landau levels are indicated with blue lines. The red arrows indicate zero-mode Landau levels. The white dashed lines indicate a measure of filling factors $\nu$. **b** A map of the Raman spectral intensity of G band of the same sample. The square shows the region relevant to the magnetotransport measurements. At any point in the square region, the Raman spectral shapes for G' band spectra were approximately the same as the AB-stacked tetralayer graphene shown in Figure S3a. **c** Landau fan diagrams, which is the same as Fig. 3**c** and **d** in the main text, are shown to compare with panel **a**.  Some energy gaps between Landau levels are indicated with blue lines. Red arrows indicate zero-mode Landau levels. Landau level structure is approximately the same as that of panel **a**.

**Figure S5 Reproducibility of $n_{tg}$ and $n_{bg}$ dependence of resistance at zero magnetic field. |**

**a** $n_{bg}$ and $n_{tg}$ dependence of resistivity measured in another sample.   **b** Derivative of resistivity with respect to $n_{bg}$. $T$ = 4.2 K. The resistance ridges arising from band structure are shown with white dashed lines. The ridge structures a b c and d were observed. Ridges e and f are newly observed structure in the hole regime that would correspond to ridges b and c in the electron regime.   **c** The same as Fig. 2b in the main text. The rectangular regime indicated by yellow lines in panel **b** corresponds to the regime for this measurement. The positions of ridges **e** and **f** are indicated.

**Figure S6 Reproducibility of $n_{tot}$ and $n_{tg}$ dependence of resistance at zero magnetic field. |**

**a** The replot of Fig. S5**a** as a function of $n_{tot}$ and $D_\perp$. **b** The replot of Fig S5**c** as a function of $n_{tot}$ and $D_\perp$.

**Figure S7 The results in AB-stacked six-layer graphene. |**

**a** Top and bottom gate voltage dependence of resistance (*R*) at zero magnetic field. Gate voltages were converted to the carrier densities $n_{tg}$ and $n_{bg}$ associated with the gate voltages. $T$ = 4.2 K.   **b** Replot as a function of $n_{tot}$ and $D_\perp$. **c** Derivative of *R* with respect to $D_\perp$.

**Figure S8 The Landau levels in tetralayer graphene. |**

**a** Landau levels in the absence of external potential variation. Results for K and K' valleys are degenerated. **b** Landau levels in the presence of external potential variation. Red and black lines are those for the K valley and the K' valley, respectively. The SWMcC parameters for these calculations were the same as those of graphite. **c** The definition of SWMcC parameters.

**Figure S9 The effect of staggered potential on Landau fan diagram |**

**a** Calculated Landau fan diagrams for different values of staggered potential amplitude. Density of states was plotted with respect to total carrier $n_{tot}$. An inset illustrates the definition of staggered potential. The position of zero-mode Landau level of light-mass bilayer-like band is shown by arrows. The zero-mode Landau levels split for $u \neq 0$. SWMcC parameters for these calculations are $\gamma_0 = 3.16$ eV, $\gamma_1 = 0.39$ eV, $\gamma_2 = -0.02$ eV, $\gamma_3 = 0.3$ eV, $\gamma_4 = 0.04$ eV, $\gamma_5 = 0.038$ eV, and $\Delta_p = 0.037$ eV. Bottom inset shows a schematic diagram of the staggered potential in each layer. **b** Experimental Landau fan diagram of AB-stacked tetra-layer graphene at $D_\perp = 0$. The width of the zero-mode Landau level is smaller than that of $u = 0.001$ eV. One can estimate that staggered potential in actual sample is less than about 0.0005 eV.

**Figure S10 Dispersion relation of tetralayer graphene with a perpendicular electric field. 1 |**

Numerically calculated dispersion relation of AB-stacked tetralayer graphene with a perpendicular electric field applied by using top and bottom gate electrodes. $n_t$ and $n_b$ are charges induced by top and bottom electrodes. SWMcC parameters were taken the same as graphite: $\gamma_0 = 3.16$ eV, $\gamma_1 = 0.39$ eV, $\gamma_2 = -0.02$ eV, $\gamma_3 = 0.3$ eV, $\gamma_4 = 0.04$ eV, $\gamma_5 = 0.038$ eV, and $\Delta_p = 0.037$ eV. Form the left to the right, $n_t$ and $n_b$ were varied as $n_t = -n_b = 0, 1 \times 10^{12}, 2 \times 10^{12}, 3 \times 10^{12}$ cm$^{-2}$. X indicates a point where the conduction and valence bands touch or overlap. Bands B and C touch near $E/\gamma_0 = 0$.

**Figure S11 Dispersion relation of tetralayer graphene with a perpendicular electric field. 2 |**

Enlargement of Fig. S10 near $E/\gamma_0 = 0$.

**Figure S12 Dispersion relation of tetralayer graphene with a perpendicular electric**

field and with smaller $\gamma_3$ |

Dispersion relation which was calculated for $\gamma_3 = 0.28$ eV. Other parameters are the same as those for Figs. S10 and S11.

**Figure S13 Dispersion relation of tetralayer graphene with a perpendicular electric field and with staggered potential 1|**

Dispersion relation which was calculated for $u = 0.5$ meV. Other parameters are the same as those for Figs. S10 and S11.

**Figure S14 Dispersion relation of tetralayer graphene with a perpendicular electric field and with staggered potential 2|**

Dispersion relation which was for $u = 1$ meV. Other parameters are the same as those for Figs. S10 and S11.

**Figure S15 Dielectric breakdown in gate voltage dependence of resistance.|**

Top and bottom gate voltage dependence of resistance in an AB-stacked six-layer graphene sample which shows a dielectric break down at about $|V_{tg}|$ =5 V. Measured sample is different from that six-layer graphene whose result of experiment is presented in Figs. S3**a** - **c**.

Fig. S1

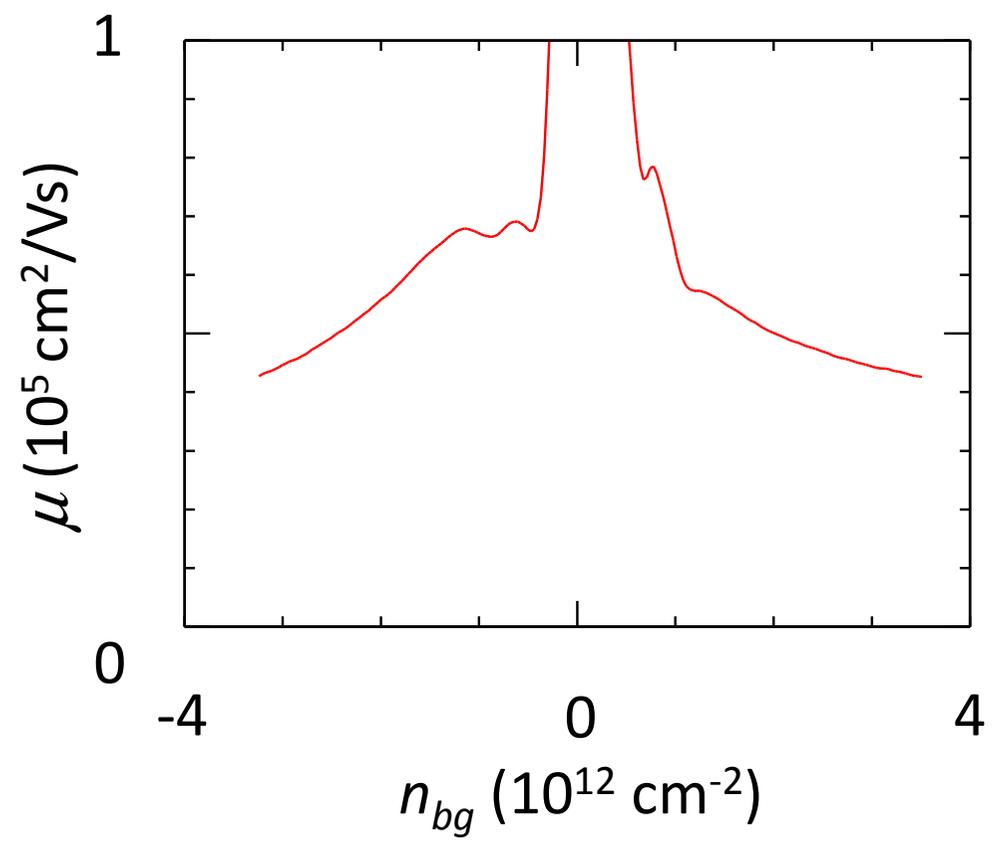

Fig. S2

a

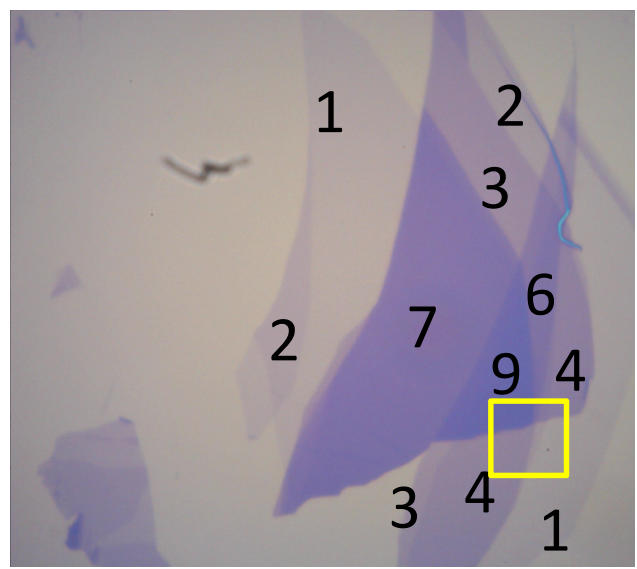

5 μm

b

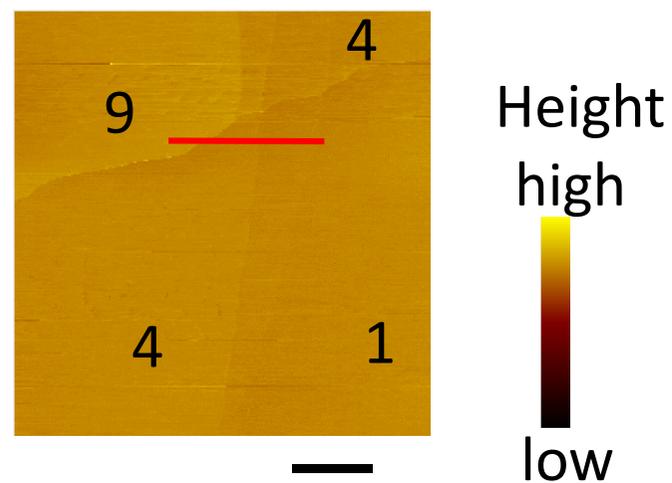

Height high

low

1 μm

c

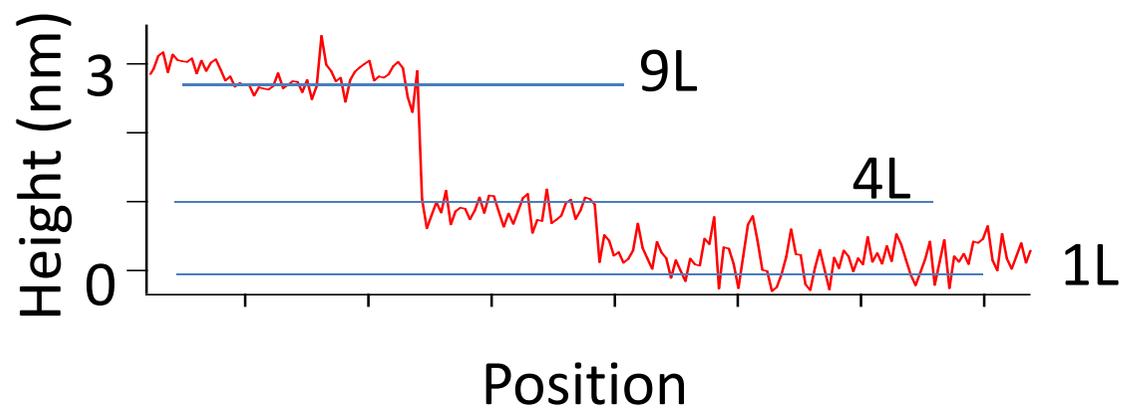

Height (nm)

9L

4L

1L

Position

Fig. S3

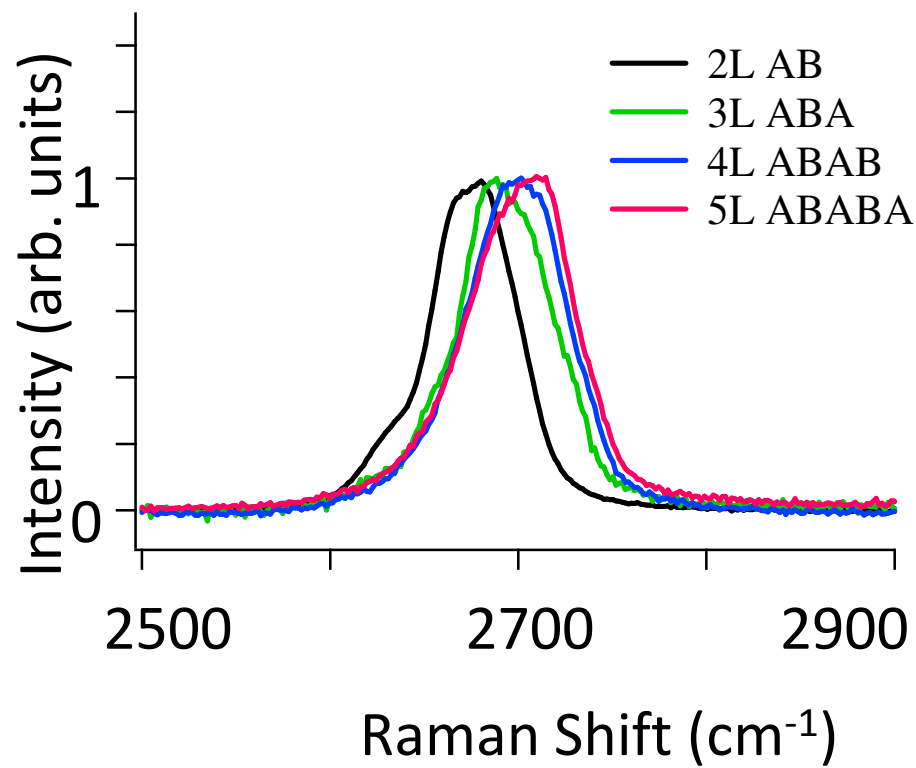 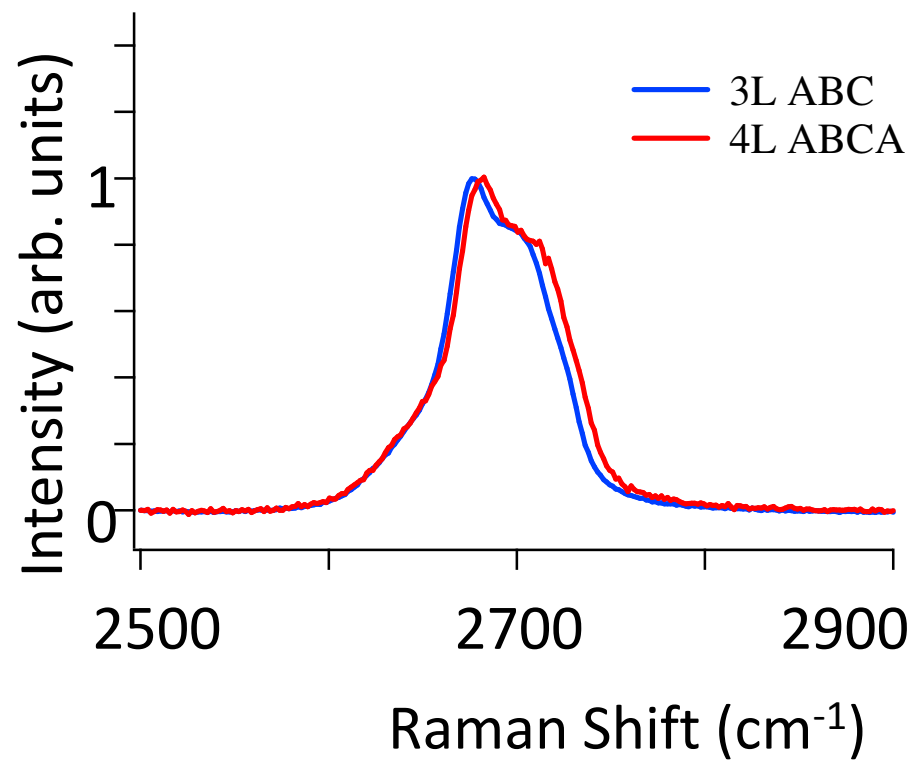

Fig. S4

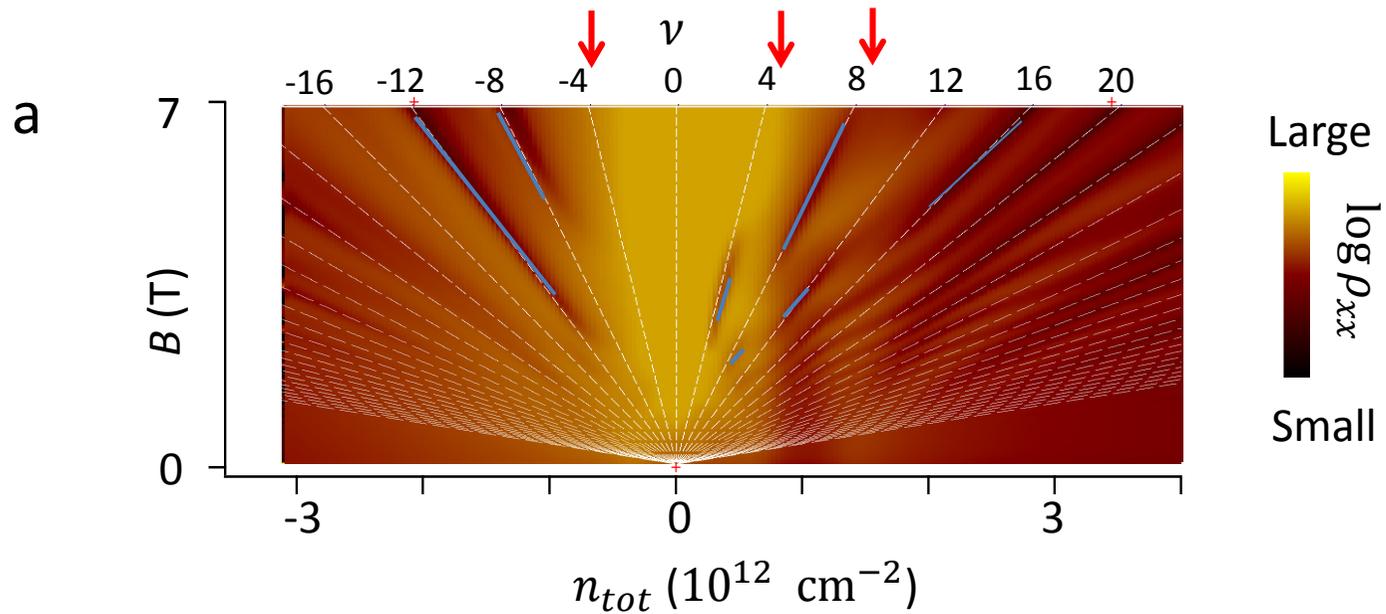

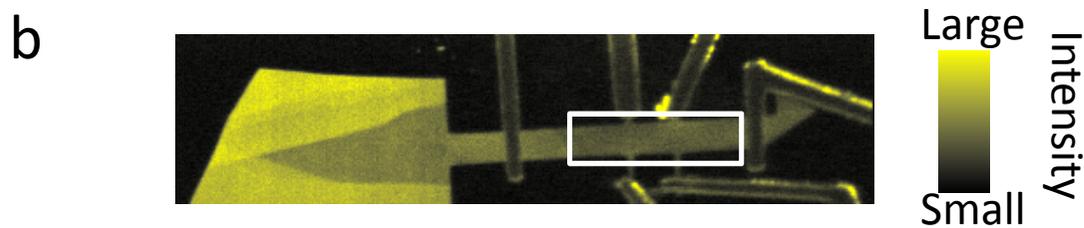

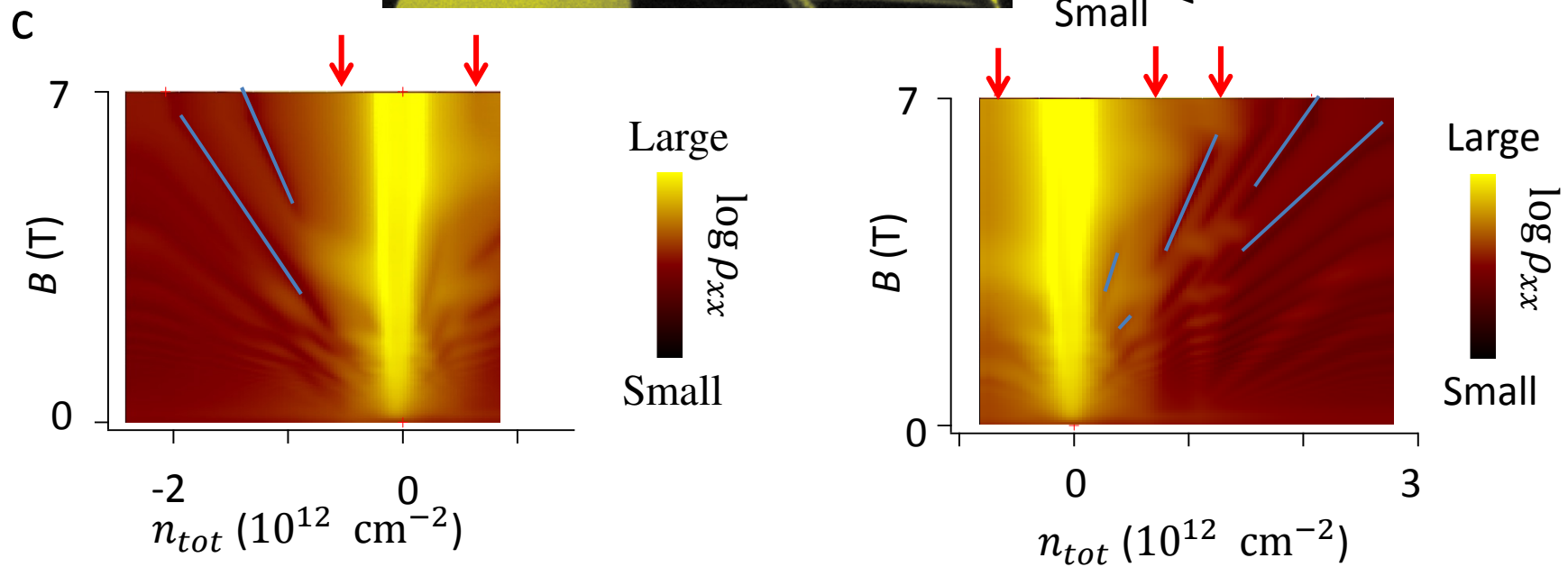

Fig. S5

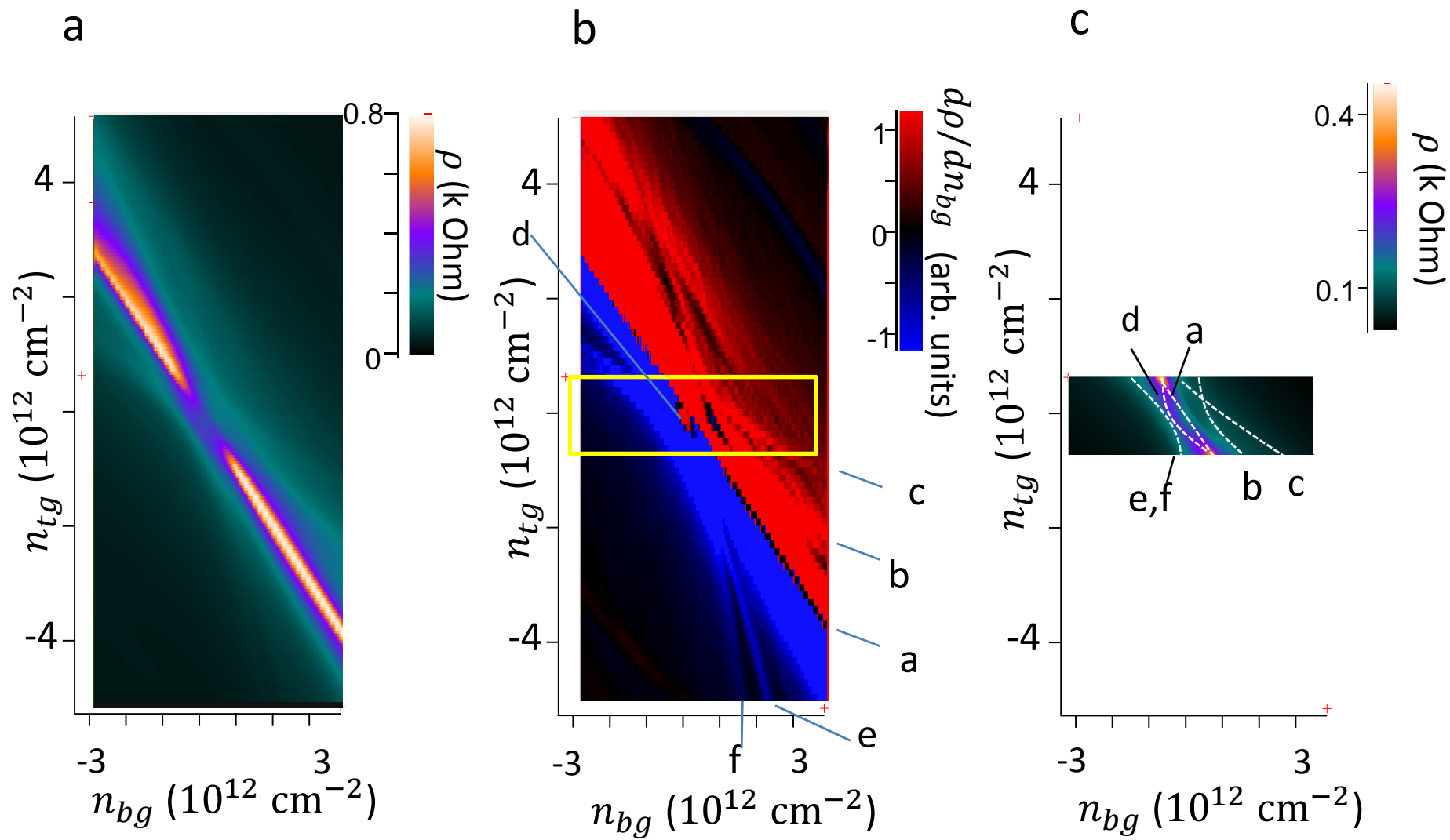

Fig. S6

a 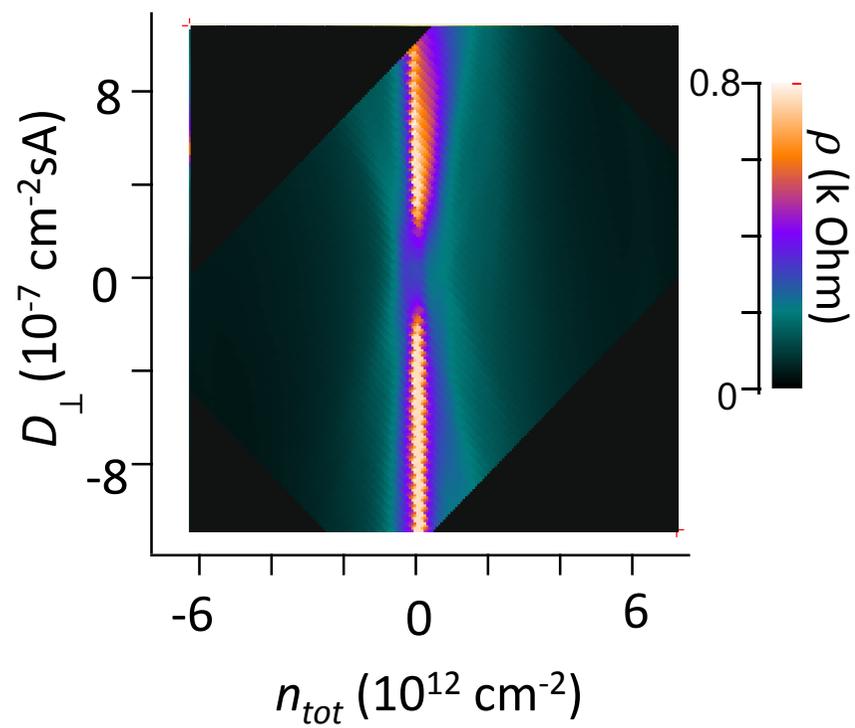
b 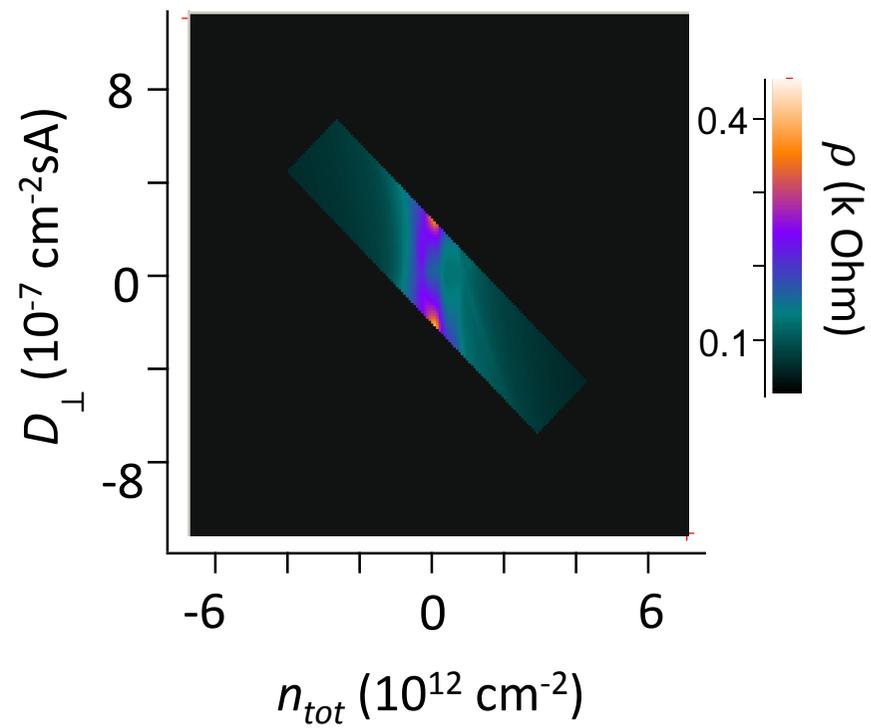

Fig. S7

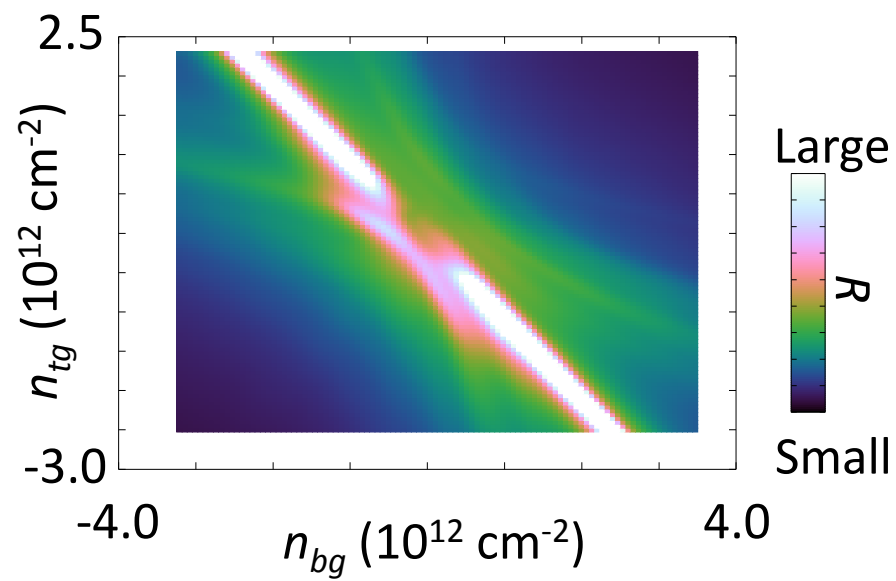
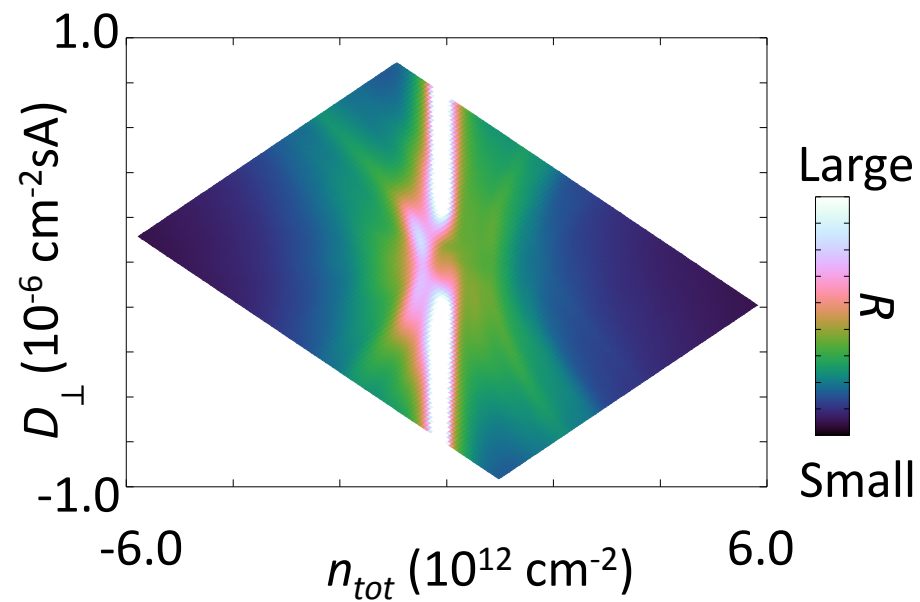
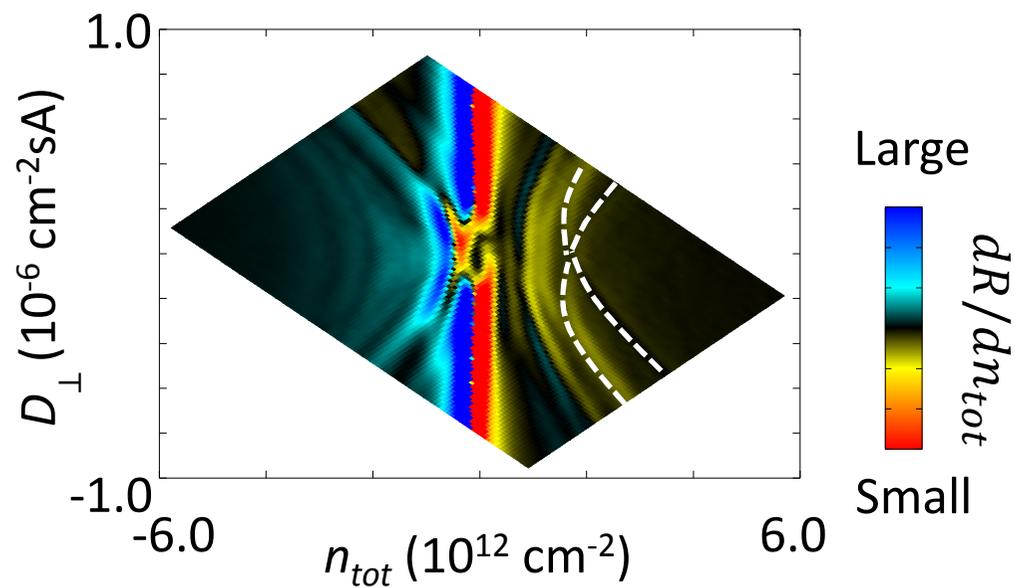

Fig. S8

a

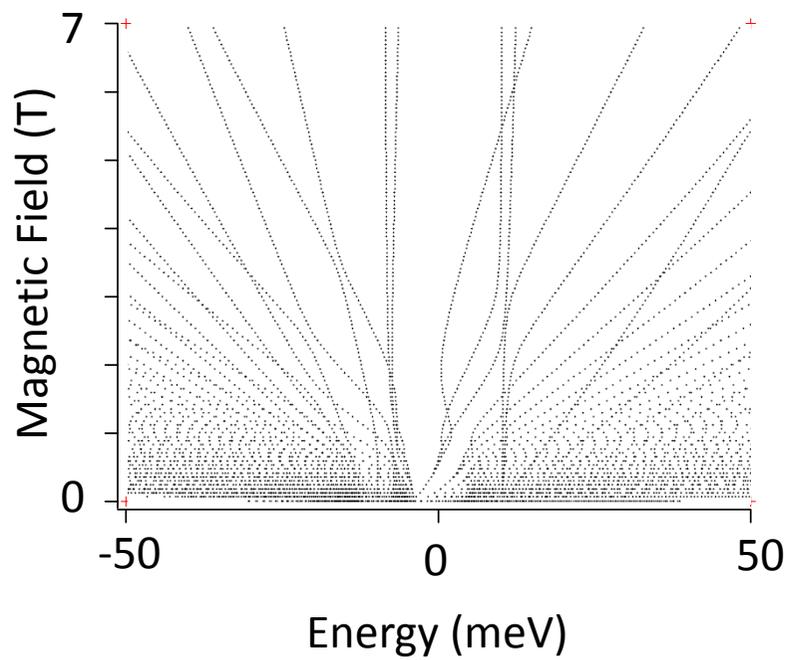

b

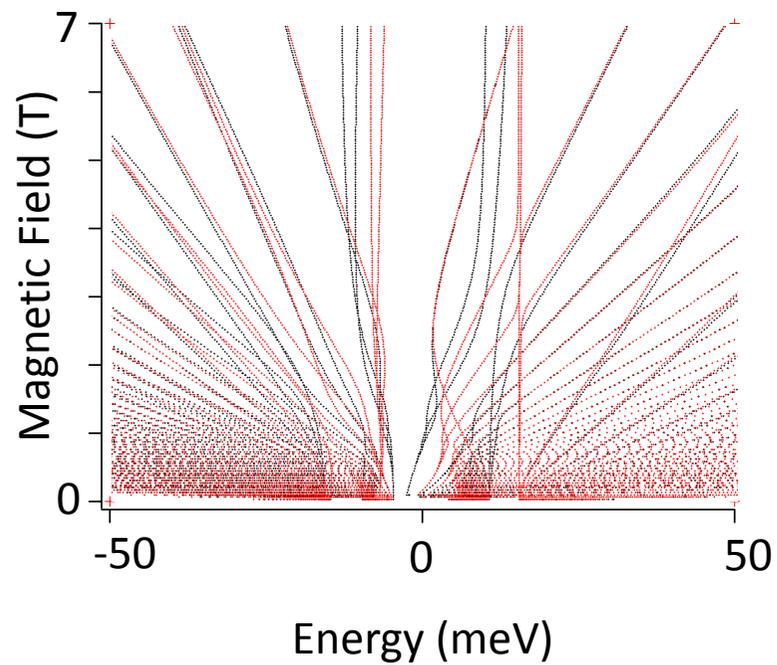

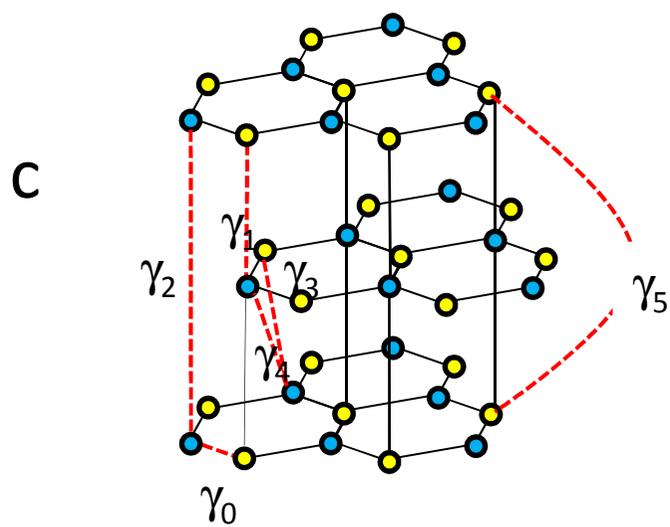

c

Fig. S9

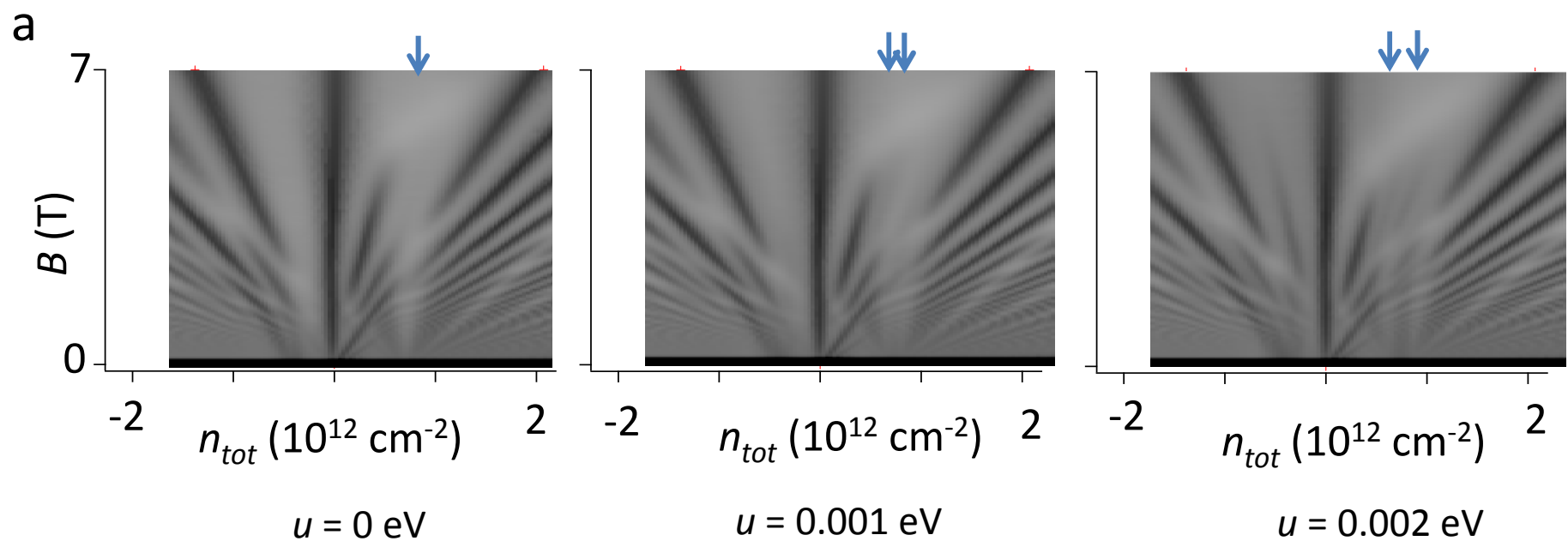

a

$u = 0$ eV  $u = 0.001$ eV  $u = 0.002$ eV

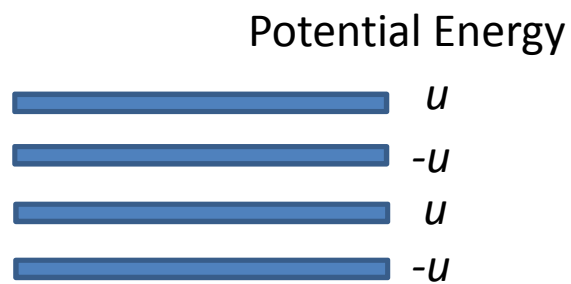

Potential Energy

$u$
$-u$
$u$
$-u$

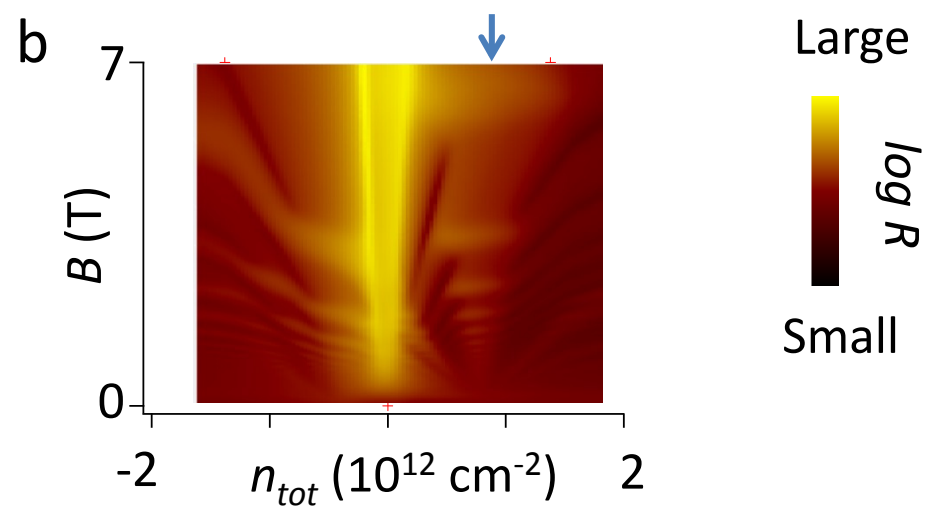

b

Fig. S10

$n_t = 0 \text{ cm}^{-2}$    $n_t = 1 \times 10^{12} \text{ cm}^{-2}$    $n_t = 2 \times 10^{12} \text{ cm}^{-2}$    $n_t = 3 \times 10^{12} \text{ cm}^{-2}$

$n_b = 0 \text{ cm}^{-2}$    $n_b = -1 \times 10^{12} \text{ cm}^{-2}$    $n_b = -2 \times 10^{12} \text{ cm}^{-2}$    $n_b = -3 \times 10^{12} \text{ cm}^{-2}$

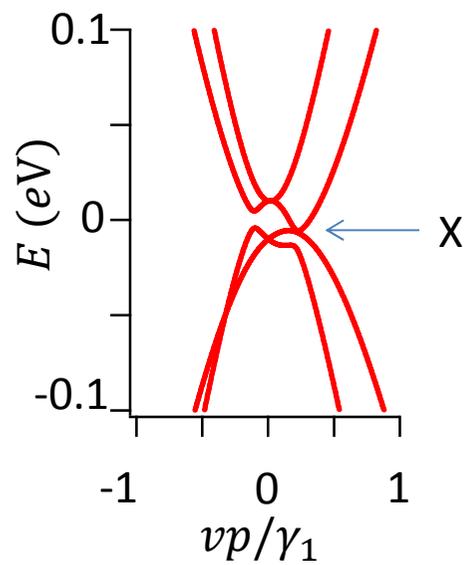
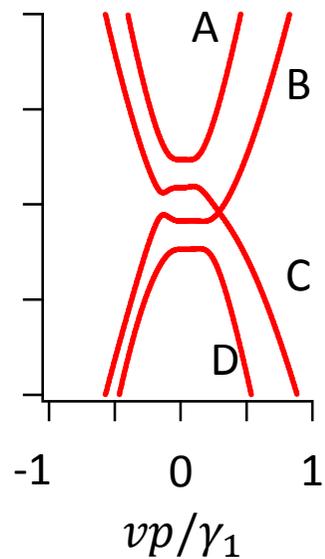
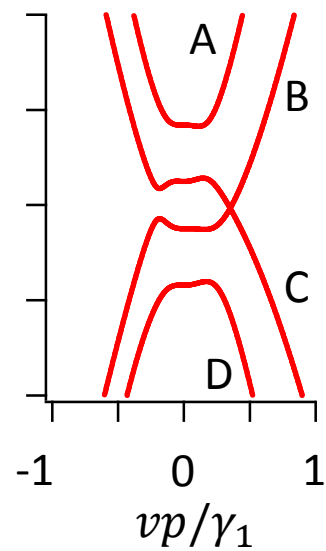
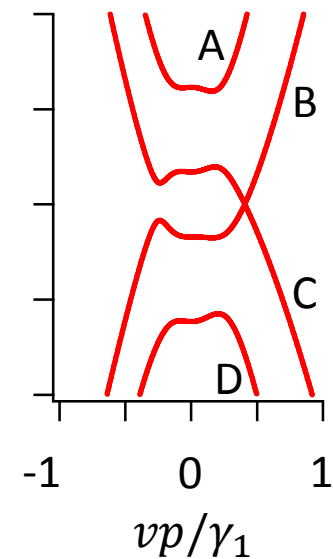

# Fig. S11

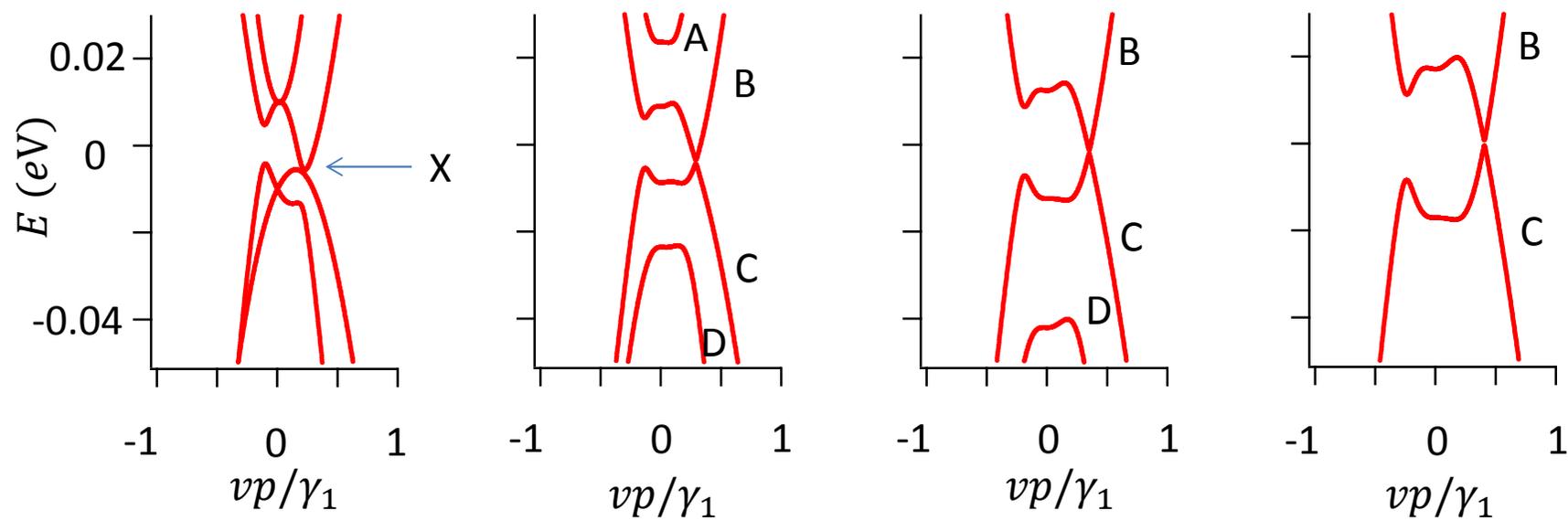

Fig. S12

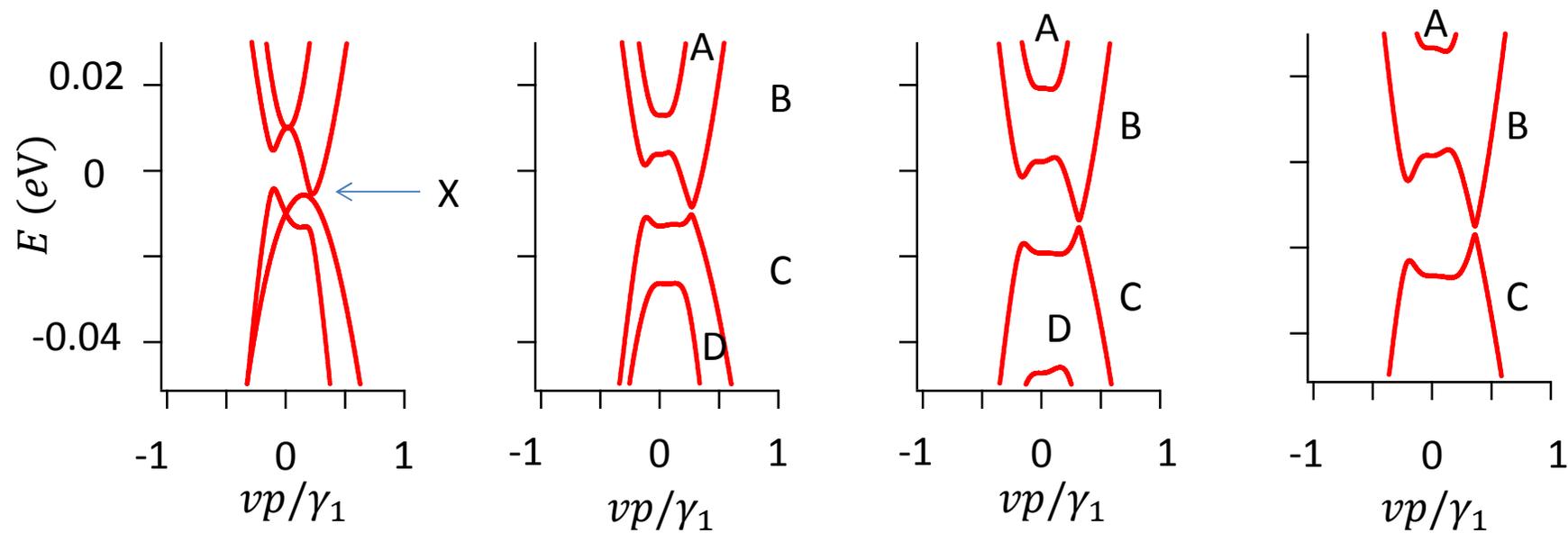

# Fig. S13

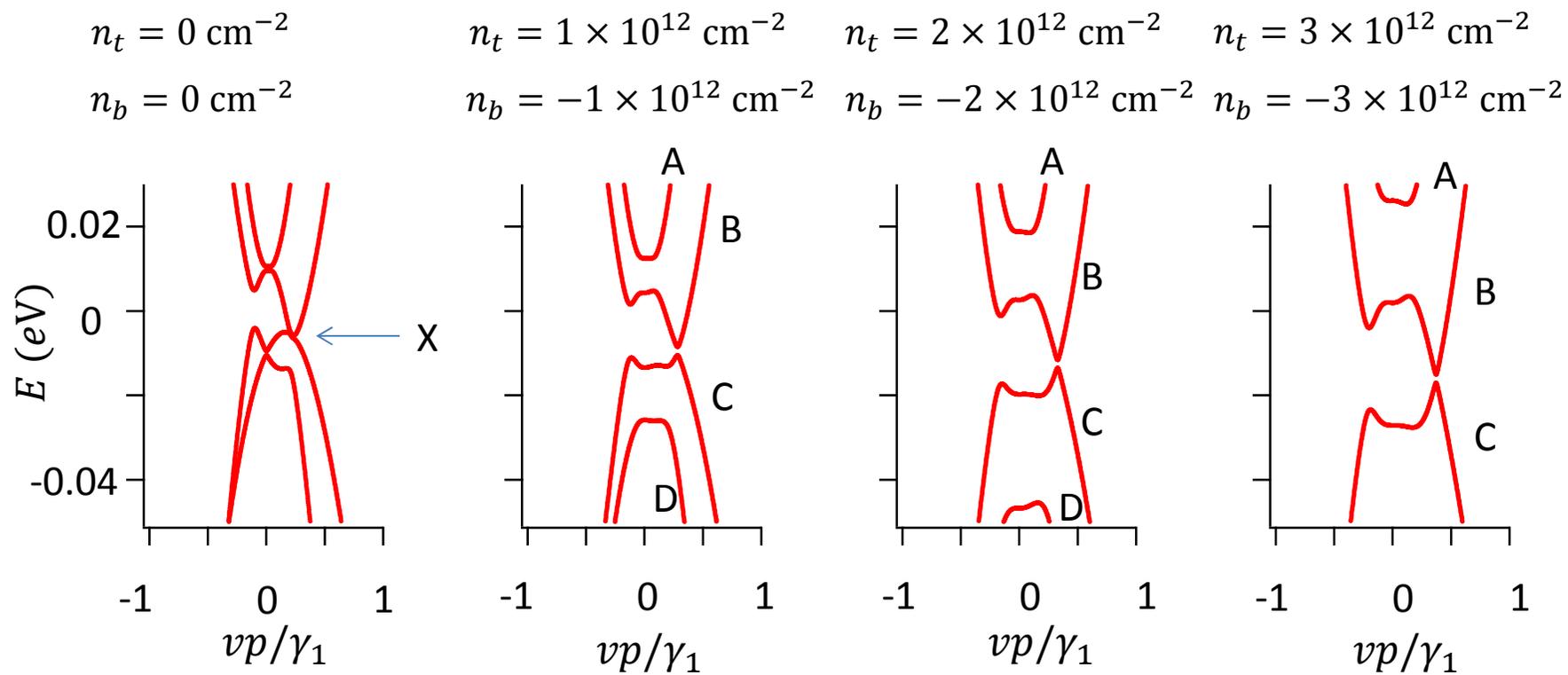

Fig. S14

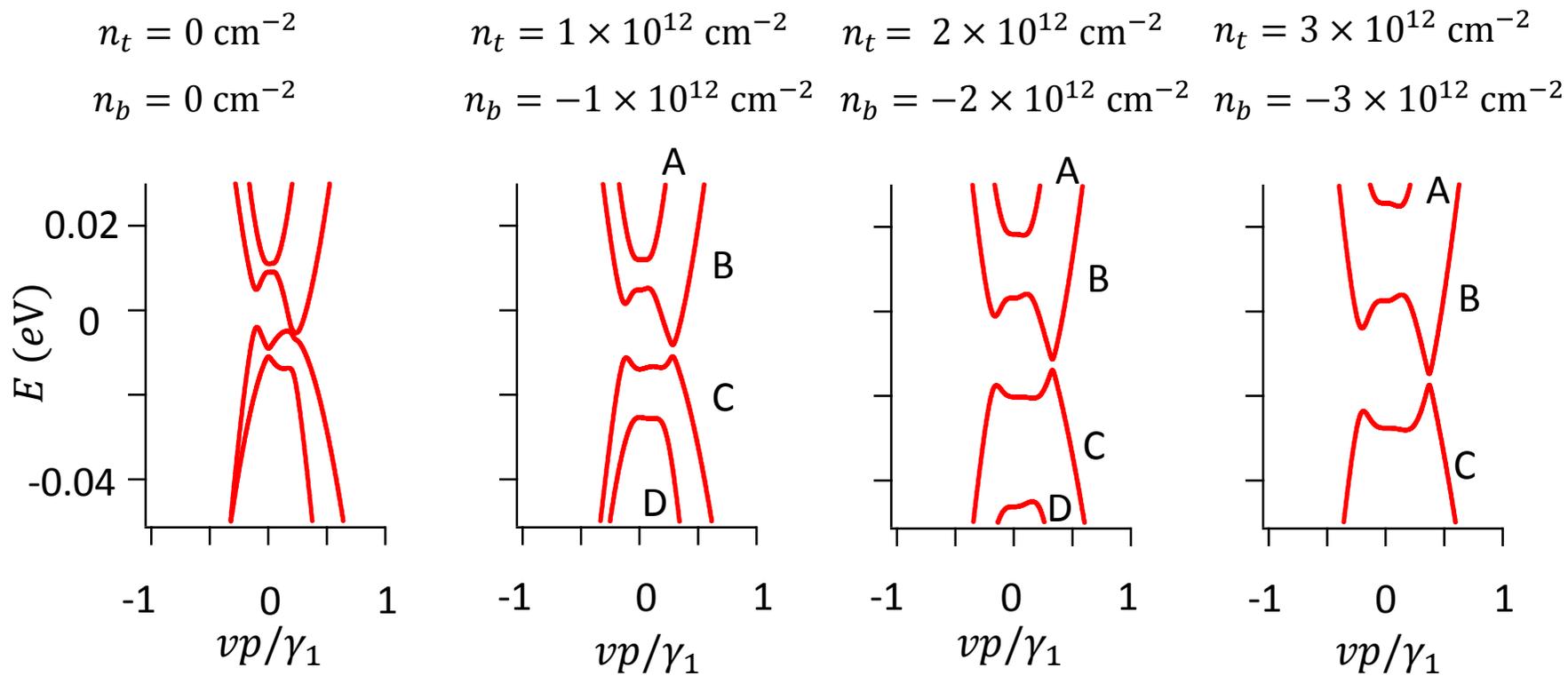

Fig. S15

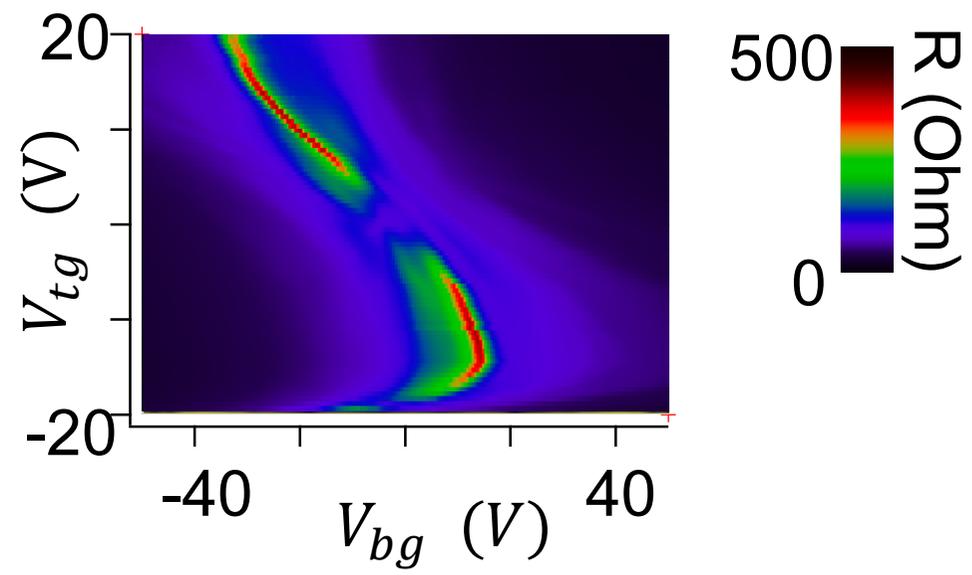